\documentclass[a4paper,11pt]{article}
\usepackage{jheppub}
\usepackage[dvipsnames]{xcolor}
\usepackage{caption}
\usepackage{subcaption}
\usepackage{xspace}
\usepackage{amsmath}
\usepackage{lineno}
\usepackage{multirow}

\usepackage{todonotes}

\newcommand{\abs}[1]{\ensuremath{\lvert #1 \rvert}}
\newcommand{\HT}{\ensuremath{H_{\mathrm{T}}}\xspace}
\newcommand{\kt}{\ensuremath{k_{\mathrm{T}}}\xspace}
\newcommand{\pt}{\ensuremath{p_{\mathrm{T}}}\xspace}

\newcommand{\GeV}{\ensuremath{\,\text{Ge\hspace{-.08em}V}}\xspace}
\newcommand{\TeV}{\ensuremath{\,\text{Te\hspace{-.08em}V}}\xspace}
\newcommand{\PQc}{\ensuremath{\mathrm{c}}\xspace}
\newcommand{\PQb}{\ensuremath{\mathrm{b}}\xspace}

\newcommand{\bbbar}{\ensuremath{\mathrm{b\overline{b}}}\xspace}
\newcommand{\PH}{\ensuremath{\mathrm{H}}\xspace}
\newcommand{\Hbb}{\ensuremath{\PH\to\bbbar}\xspace}
\newcommand{\HH}{\ensuremath{\mathrm{HH}}\xspace}
\newcommand{\HHH}{\ensuremath{\mathrm{HHH}}\xspace}
\newcommand{\MADGRAPH} {\textsc{MadGraph}\xspace}
\newcommand{\MCATNLO} {\textsc{mc@nlo}\xspace}
\newcommand{\PYTHIA} {{\textsc{pythia}}\xspace}
\newcommand{\MGvATNLO}{\MADGRAPH{}5\_a\MCATNLO}
\newcommand{\DELPHES} {{\textsc{delphes}}\xspace}
\newcommand{\DP}{\ensuremath{\mathrm{DP}}\xspace}
\newcommand{\AP}{\ensuremath{\mathrm{AP}}\xspace}
\newcommand{\SD}{\ensuremath{\mathrm{SD}}\xspace}
\DeclareMathOperator*{\argmax}{arg\,max}

\title{Reconstruction of boosted and resolved multi-Higgs-boson events with symmetry-preserving attention networks}

\author[a]{Haoyang Li,}
\author[b]{Marko Stamenkovic,}
\author[c]{Alexander Shmakov,}
\author[c]{Michael Fenton,}
\author[a]{Darius Shih-Chieh Chao,}
\author[a]{Kaitlyn Maiya White,}
\author[d]{Caden Mikkelsen,}
\author[e]{Jovan Miti\'{c},}
\author[f]{Cristina Mantilla Suarez,}
\author[a]{Melissa Quinnan,}
\author[b]{Greg Landsberg,}
\author[d]{Harvey Newman,}
\author[c]{Pierre Baldi,}
\author[c]{Daniel Whiteson,}
\author[a]{and Javier Duarte}
\affiliation[a]{UC San Diego, La Jolla, CA, USA}
\affiliation[b]{Brown University, Providence, RI, USA}
\affiliation[c]{UC Irvine, Irvine, CA, USA}
\affiliation[d]{Caltech, Pasadena, CA, USA}
\affiliation[e]{University of Belgrade, Belgrade, Serbia}
\affiliation[f]{University of Virginia, VA, USA}

\emailAdd{marko\_stamenkovic@brown.edu}

\abstract{
The production of multiple Higgs bosons at the CERN LHC provides a direct way to measure the trilinear and quartic Higgs self-interaction strengths as well as potential access to beyond the standard model effects that can enhance production at large transverse momentum \pt.
The largest event fraction arises from the fully hadronic final state in which every Higgs boson decays to a bottom quark-antiquark pair (\bbbar). 
This introduces a combinatorial challenge known as the \emph{jet assignment problem}: assigning jets to sets representing Higgs boson candidates.
Symmetry-preserving attention networks (SPA-Nets) have been been developed to address this challenge.
However, the complexity of jet assignment increases when simultaneously considering both \Hbb reconstruction possibilities, i.e., two ``resolved'' small-radius jets each containing a shower initiated by a \PQb quark or one ``boosted'' large-radius jet containing a merged shower initiated by a \bbbar pair.
The latter improves the reconstruction efficiency at high \pt.
In this work, we introduce a generalization to the SPA-Net approach to simultaneously consider both boosted and resolved reconstruction possibilities and unambiguously interpret an event as ``fully resolved,'' ``fully boosted,'' or in between.
We report the performance of baseline methods, the original SPA-Net approach, and our generalized version on nonresonant \HH and \HHH production at the LHC.
Considering both boosted and resolved topologies, our SPA-Net approach increases the Higgs boson reconstruction purity by 56--80\% and the efficiency by 37--38\% compared to the baseline method depending on the final state.
}

\begin{document}
\maketitle
\flushbottom

\section{Introduction}
\label{sec:intro}

\begin{figure}[htpb]
    \centering
    \includegraphics[width=0.49\textwidth]{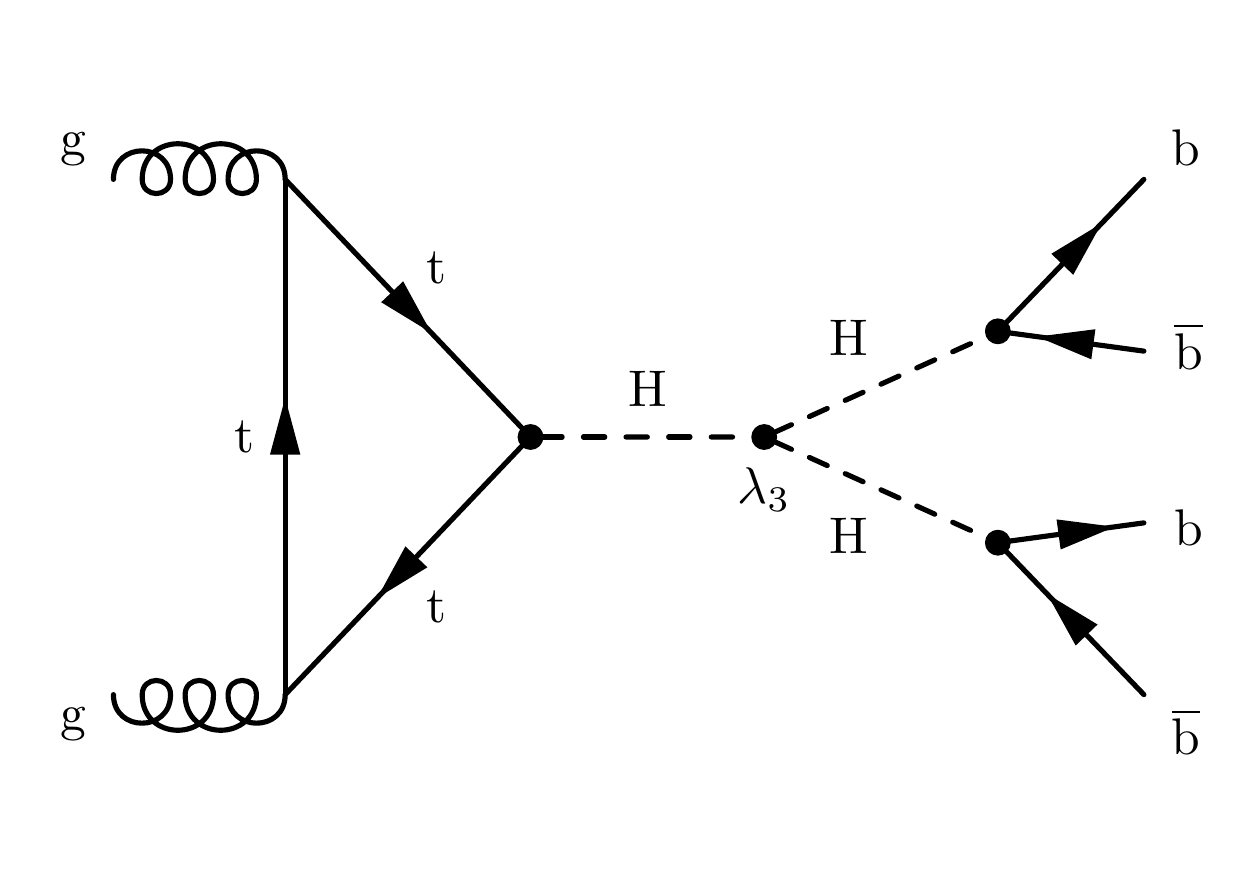}
    \includegraphics[width=0.49\textwidth]{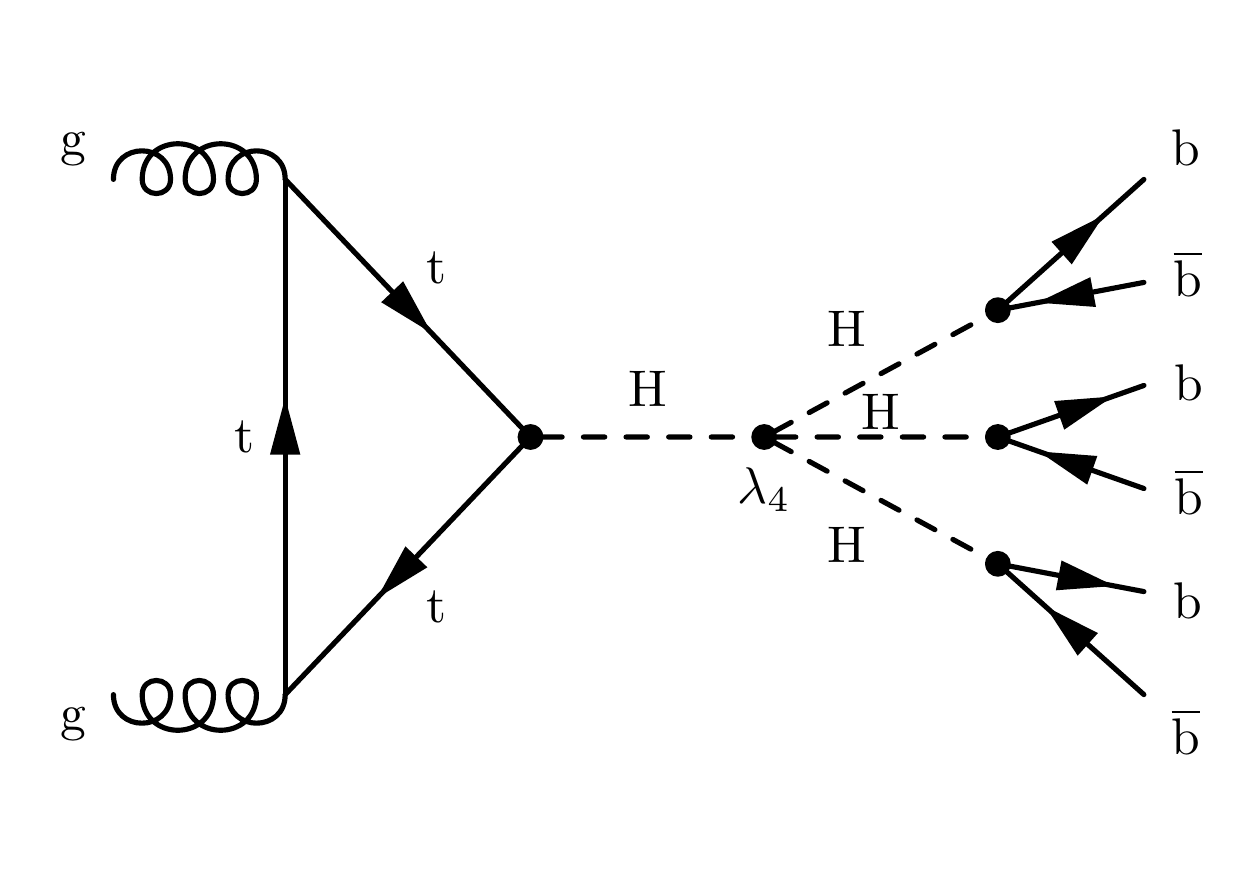}
    \caption{Representative Feynman diagrams for the production of two Higgs bosons (left) and three Higgs bosons (right) in the gluon-gluon fusion process, where all Higgs bosons decay to \bbbar pairs.}
    \label{fig:feynman}
\end{figure}

Measuring multiple Higgs boson production at the LHC is a powerful probe of the Higgs trilinear ($\lambda_3$) and quartic ($\lambda_4$) self-couplings as well as new physics effects beyond the standard model (SM), such as resonant production via a heavy particle decaying into two Higgs bosons.
While the quartic Higgs self-coupling is notoriously difficult to measure at the LHC, current theoretical studies indicate that even at a future 100 \TeV\ hadron collider the Standard Model \HHH\ production process would be observable with only a 2–3~$\sigma$ significance~\cite{Cepeda:2019klc,Plehn:2005nk,Papaefstathiou:2019ofh,Fuks:2017zkg,Contino:2016spe}. Nevertheless, triple Higgs production remains a direct probe of this coupling as well as physics beyond the SM and has recently attracted increasing attention in LHC studies~
\cite{Brigljevic:2024vuv}. By contrast, the trilinear self-coupling can be accessed more directly through measurements of Higgs boson pair production~\cite{DiMicco:2019ngk}.
These processes can also exhibit modified kinematics and enhanced cross sections at large transverse momentum (\pt) for non-SM coupling values~\cite{Dolan:2013rja,Dolan:2015zja,Bishara:2016kjn,Carvalho_2016,Arganda:2018ftn,Capozi_2020}, which makes it important to accurately reconstruct and measure them across a wide range of \pt.
The dominant decay mode of the Higgs boson is into a bottom quark-antiquark (\bbbar) pair, which leads to a fully hadronic final state with multiple jets.
Representative Feynman diagrams for the production of two or three Higgs bosons in the gluon-gluon fusion process where all Higgs bosons decay to \bbbar pairs are shown in Fig.~\ref{fig:feynman}.
A key challenge in this final state is to find the optimal association of jets to Higgs boson candidates.
This problem is combinatorial in nature and becomes computationally challenging for conventional methods as the number of jets increases.

\begin{figure}[htpb]
    \centering
    \includegraphics[height=0.3\linewidth]{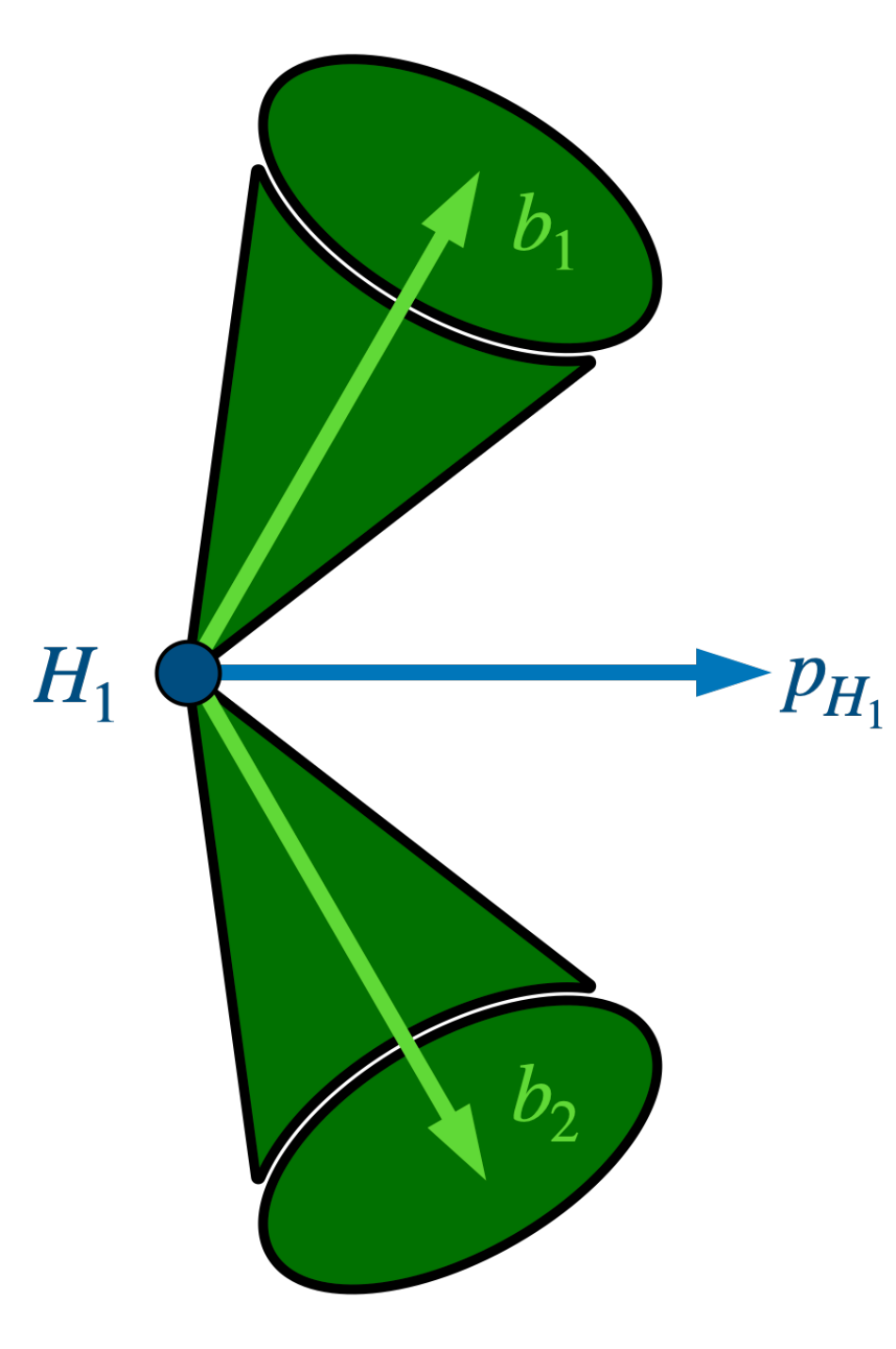}
    \includegraphics[height=0.3\linewidth]{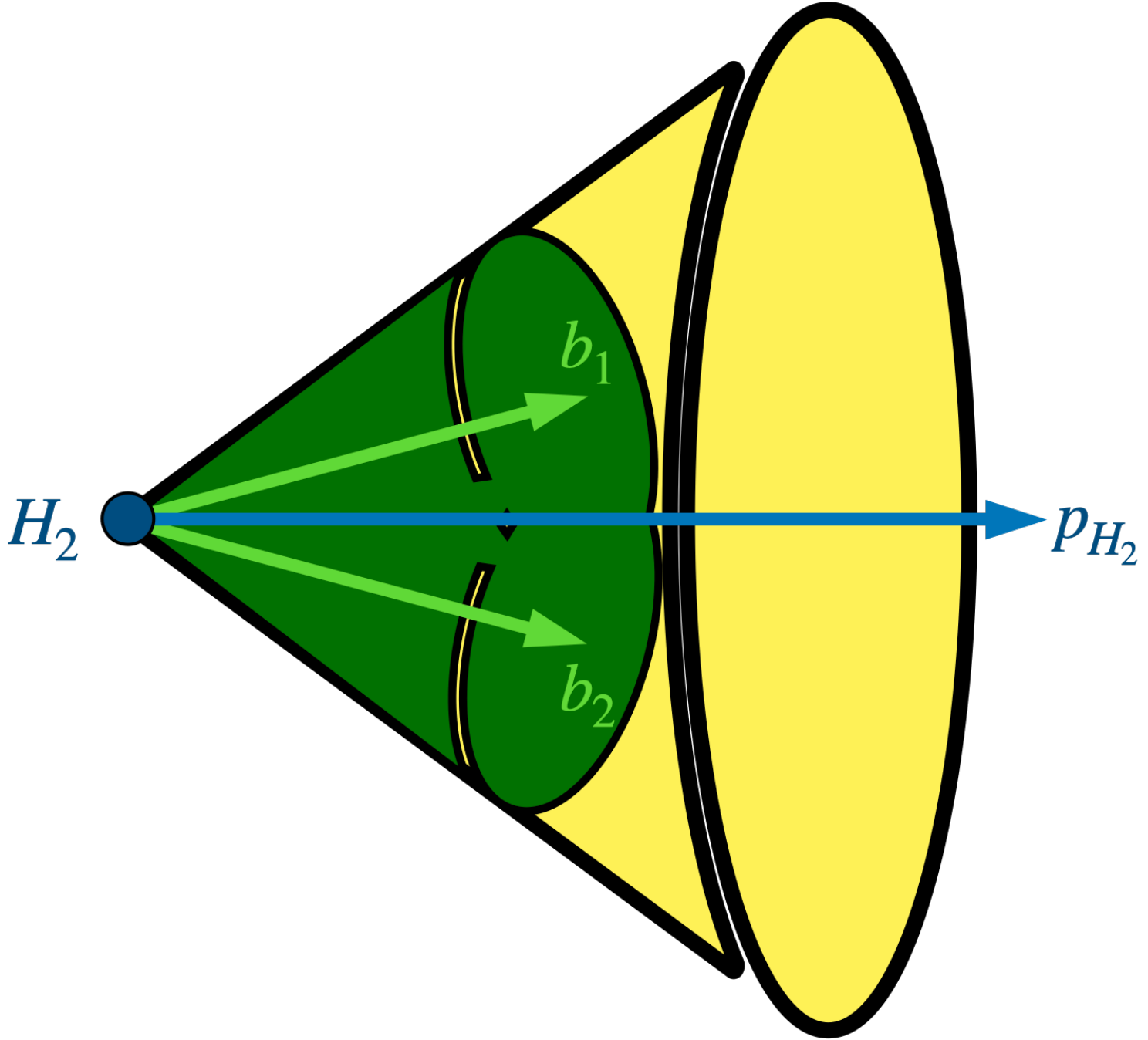}
    \includegraphics[height=0.3\linewidth]{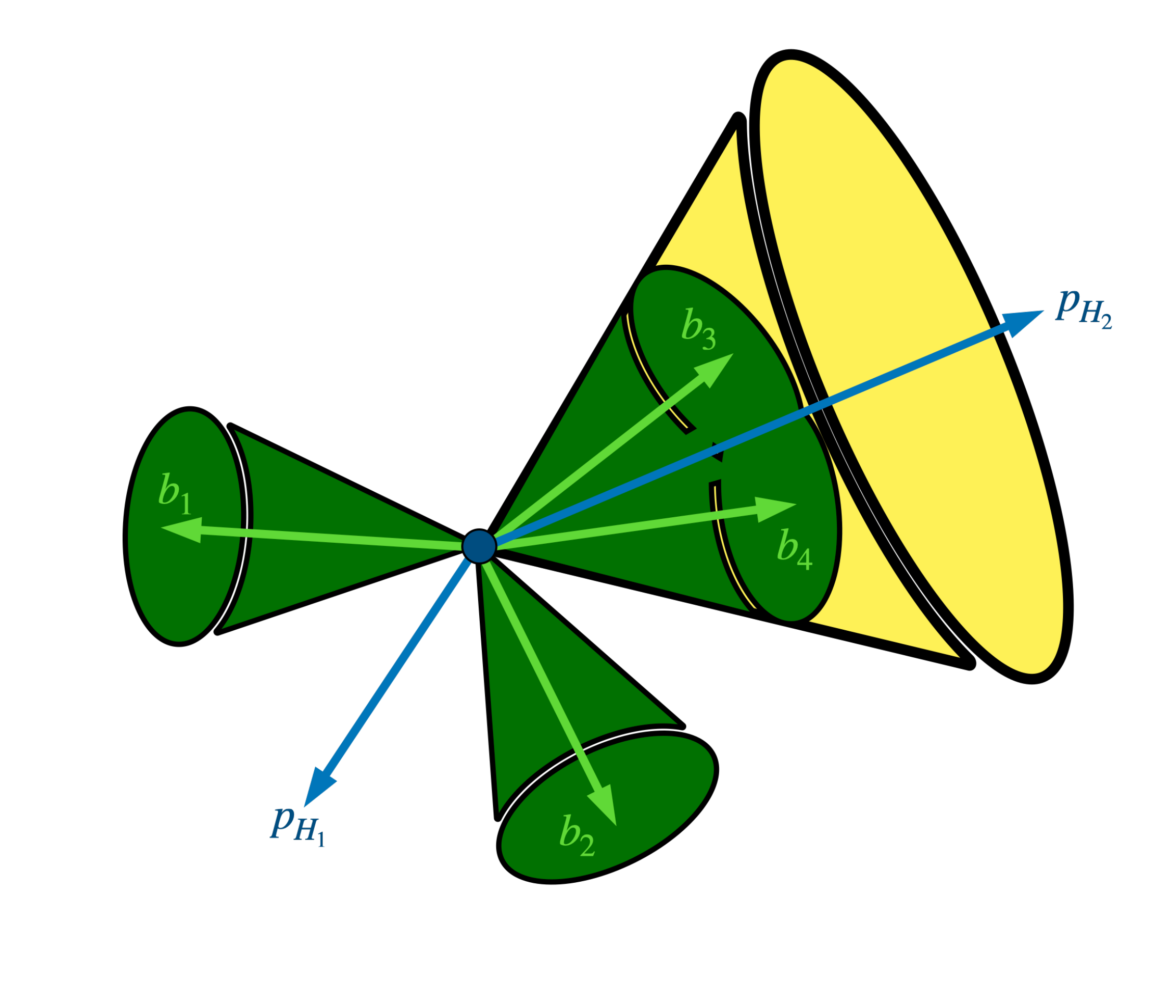}
    \caption{Two bottom quarks from a Higgs boson decay are reconstructed as two separated small-$R$ jets in the resolved topology when the Higgs boson $\pt \lesssim 2m_\PH/R$ (left).
    Here, $m_\PH$ denotes the Higgs boson mass.
    Two bottom quarks from a Higgs boson decay overlap and merge into a single large-$R$ jet in the boosted topology when the Higgs boson $\pt \gtrsim 2m_\PH/R$ (center).
    An event with two Higgs bosons where one is resolved and one is boosted (right)}
    \label{fig:boosted_resolved}
\end{figure}

Moreover, the complexity of this problem is further increased by the presence of different event topologies, depending on the \pt of the Higgs bosons.
At low \pt, each Higgs boson can be reconstructed as two ``resolved" small-radius jets, each containing a \PQb-quark-initiated jet.
At high \pt, each Higgs boson can be reconstructed from one ``boosted" large-radius jet, containing a merged \bbbar pair.
For intermediate \pt, a mixture of resolved and boosted jets can be present.
This is illustrated in Fig.~\ref{fig:boosted_resolved}.
The CMS and ATLAS Collaborations have performed many searches for resonant and nonresonant Higgs boson pair production in boosted~\cite{CMS:2018vjd,CMS:2022gjd} and resolved~\cite{CMS:2018qmt,CMS:2018sxu,ATLAS:2020jgy,CMS:2022cpr} final states, and their statistical combination~\cite{ATLAS:2015zug,ATLAS:2016paq,ATLAS:2018rnh,CMS:2022dwd,ATLAS:2022hwc,ATLASCombination2024,CMS-PAS-HIG-20-011}.
In CMS, the boosted category is found to be 40\% more sensitive than the resolved, illustrating the importance of having an inclusive reconstruction targeting the various topologies.
As of now, there is no experimental result targeting an intermediate event topology reconstruction.
In the context of \HH and \HHH, maximizing the signal acceptance by targeting the different topologies is important in order to enhance the sensitivity to the SM cross section.

The combination of the boosted and resolved channels is challenging because the two data samples need to be statistically independent, but the two selection criteria can often select the same events, especially for Higgs bosons in the intermediate \pt range.
Thus, analysts must implement an ``overlap removal'' procedure to allow these two searches to be combined.
Part of the motivation of this work is to use machine learning to help interpret individual events as boosted, resolved, or in some intermediate configuration.

Symmetry-preserving attention networks (SPA-Nets) are a novel class of neural networks that have been proposed~\cite{Shmakov:2021qdz,Fenton:2020woz,Fenton:2023ikr} to solve the jet assignment problem for a fixed event topology.
They have also been studied in the context of resolved nonresonant \HH events to improve the sensitivity to $\lambda_3$ at the high-luminosity LHC~\cite{Chiang:2024pho}.
In this work, we present a generalization of the SPA-Net approach to handle variable (boosted and resolved) event topologies and simultaneously consider both small- and large-radius jets.
We apply this approach to nonresonant \HH and \HHH production at the LHC and compare its performance with baseline methods and the original fully resolved SPA-Net approach.
Finally, we also demonstrate how SPA-Net can provide a partitioning of events into disjoint categories based on the number of reconstructed boosted Higgs boson candidates.

The rest of this paper is organized as follows.
Section~\ref{sec:related} discusses related work.
Section~\ref{sec:dataset} presents the dataset and ground truth labeling.
We describe the baseline methods and SPA-Net model configurations in Section~\ref{sec:methods}.
In Section~\ref{sec:postprocessing}, we delineate how we use the SPA-Net outputs to assign jets and categorize events.
We present the results in Section~\ref{sec:results} and summarize in Section~\ref{sec:summary}.

The dataset~\cite{duarte_2024_14257266} and code~\cite{javier_duarte_2024_14248961} for analyzing it are publicly available.
For implementing the training, we use the \textsc{SPANet} library version 2.3, which features new additions related to reweighting and an improved validation accuracy metric as described in Section~\ref{sec:methods}~\cite{spanet}.

\subsection{Related Work}
\label{sec:related}
Recently, the ATLAS Collaboration has also searched for resonant and nonresonant \HHH production in the $\bbbar\bbbar\bbbar$ final state~\cite{ATLAS:2024xcs}, which finds sensitivity at the level of $750\times SM$ at 95\% confidence level.
In the context of the search for $\HH\to4\PQb$, resolved, intermediate, and boosted topologies were studied in Ref.~\cite{Amacker:2020bmn} where the categorization is defined based on the presence of a large-radius jet.
In order to enhance the sensitivity to the \HH process as well as to enable inclusive approaches for searches for the \HHH process at the LHC, we propose a generalization of the jet assignment and event categorization in this paper.
Ref.~\cite{CMS:2022gjd} uses a state-of-the-art jet tagger, ParticleNet~\cite{Qu:2019gqs} in order to identify the boosted $\Hbb$ large-radius jet candidates in the $\HH\to4\PQb$ search.
Such methods can be used in combination with the method we propose, for example as a training input, making them complimentary.
As mentioned above, Ref.~\cite{Chiang:2024pho} applies the standard SPA-Net method with small-radius jets to the resolved nonresonant $\HH\to4\PQb$ search and quantifies the potential improvement in the sensitivity to $\lambda_3$ at the HL-LHC.
While our study does not aim to estimate the full analysis sensitivity as this would require background modeling, systematic uncertainties, and trigger-level considerations that are beyond the scope of our work, we aim to evaluate the improvements in signal reconstruction performance from our proposed enhancement to SPA-Net.

\section{Dataset and Labeling}
\label{sec:dataset}

\begin{figure}[htpb]
    \centering
    \includegraphics[width=0.49\linewidth]{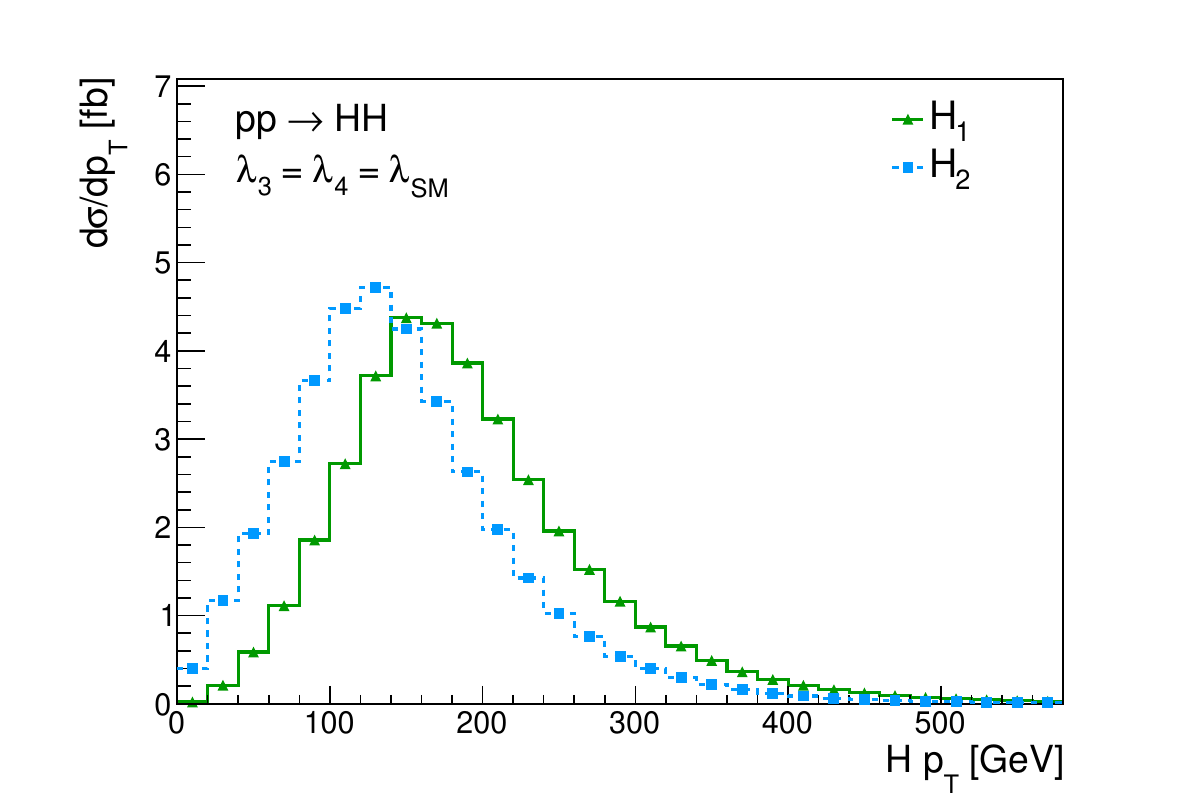}
    \includegraphics[width=0.49\linewidth]{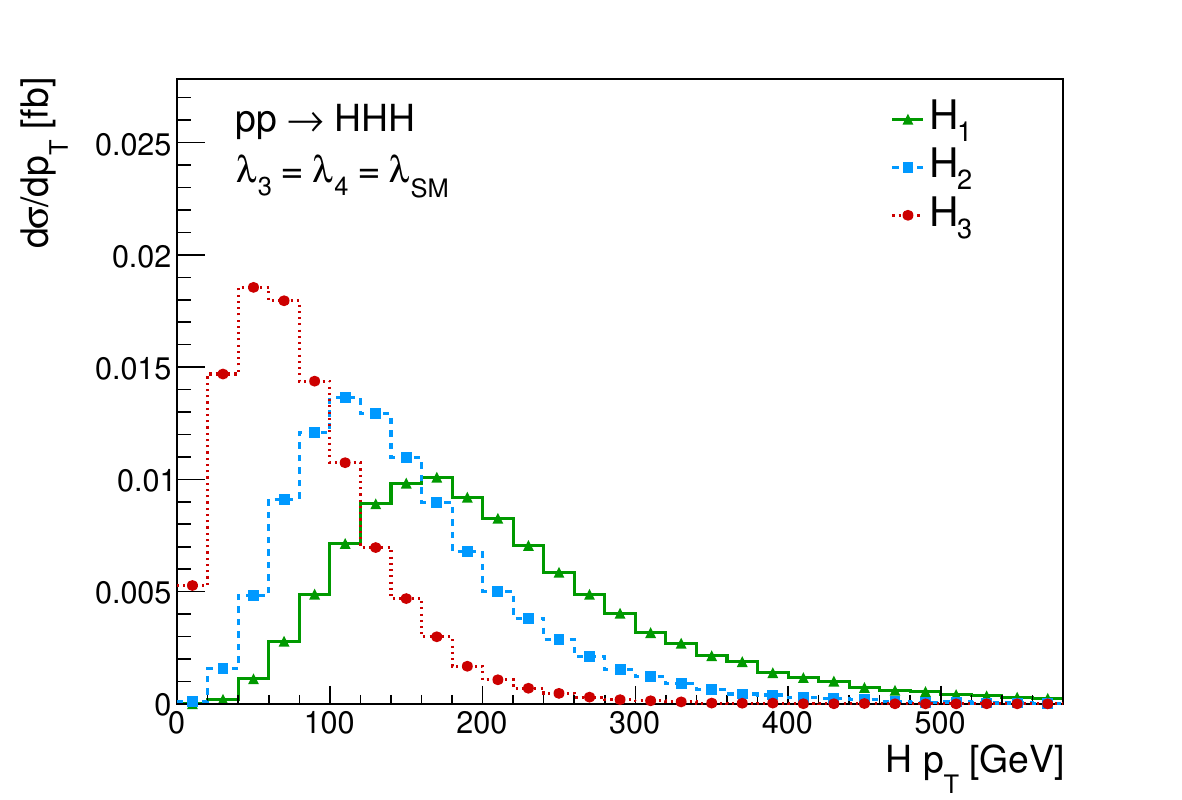}
    \caption{\HH (left) and \HHH (right) differential cross sections scaled to next-to-next-to-LO (NNLO) as a function of the individual \PH \pt observables. 
    The Higgs bosons are sorted by \pt.}
    \label{fig:hh_hhh_pt}
\end{figure}

Nonresonant \HH and \HHH events are generated using {\MGvATNLO}3.4.1~\cite{Alwall:2014hca} at a leading-order (LO) accuracy at the LHC with a center of mass energy $\sqrt{s}=14\TeV$, with the trilinear and quartic couplings fixed to the SM values as implemented in the model of Ref.~\cite{Papaefstathiou:2020lyp}.
The corresponding \HH and \HHH differential cross sections as a function of the individual \PH \pt observables is shown in Fig.~\ref{fig:hh_hhh_pt}.
Only the decay of the Higgs boson to bottom quarks is considered.
The parton shower and hadronization are simulated with {\PYTHIA}8.2~\cite{Sjostrand:2014zea} and the detector response is simulated with {\DELPHES}3.4.1~\cite{deFavereau:2013fsa}, using the CMS parametrization.

Final-state particles are clustered into jets using the anti-$\kt$ algorithm~\cite{Cacciari:2008gp,Cacciari:2011ma} with radius parameters of 0.5 (AK5 jets) and 0.8 (AK8 jets).
We require $\pt > 20\GeV$ for AK5 jets and $\pt > 200\GeV$ for AK8 jets, as well as $\abs{\eta} < 2.5$ for both.
Emulation of a \PQb-tagging algorithm is applied to the AK5 jets, which assigns a boolean value to each jet indicating whether it originates from a \PQb quark, based on \pt-dependent efficiency and misidentification rates.
The misidentification rate for light (\PQc) jets is about 1\% (20\%), while the \PQb-tagging efficiency is about 70\%~\cite{CMS:2012feb}.
We note that these are conservative relative to current state-of-the-art \PQb-tagging algorithms~\cite{Mondal:2024nsa}.
The four-momentum $(\pt, \eta, \phi, m)$ and the \PQb-tagging value of each AK5 jet are stored as inputs to the networks.
For AK8 jets, we consider the four-momentum $(\pt, \eta, \phi, m)$ as well.
No \PQb-tagging is applied for AK8 jets, however, we develop our own tagger discussed in Section~\ref{sec:baseline}.

\begin{figure}[htpb]
    \centering
    \includegraphics[width=0.7\textwidth]{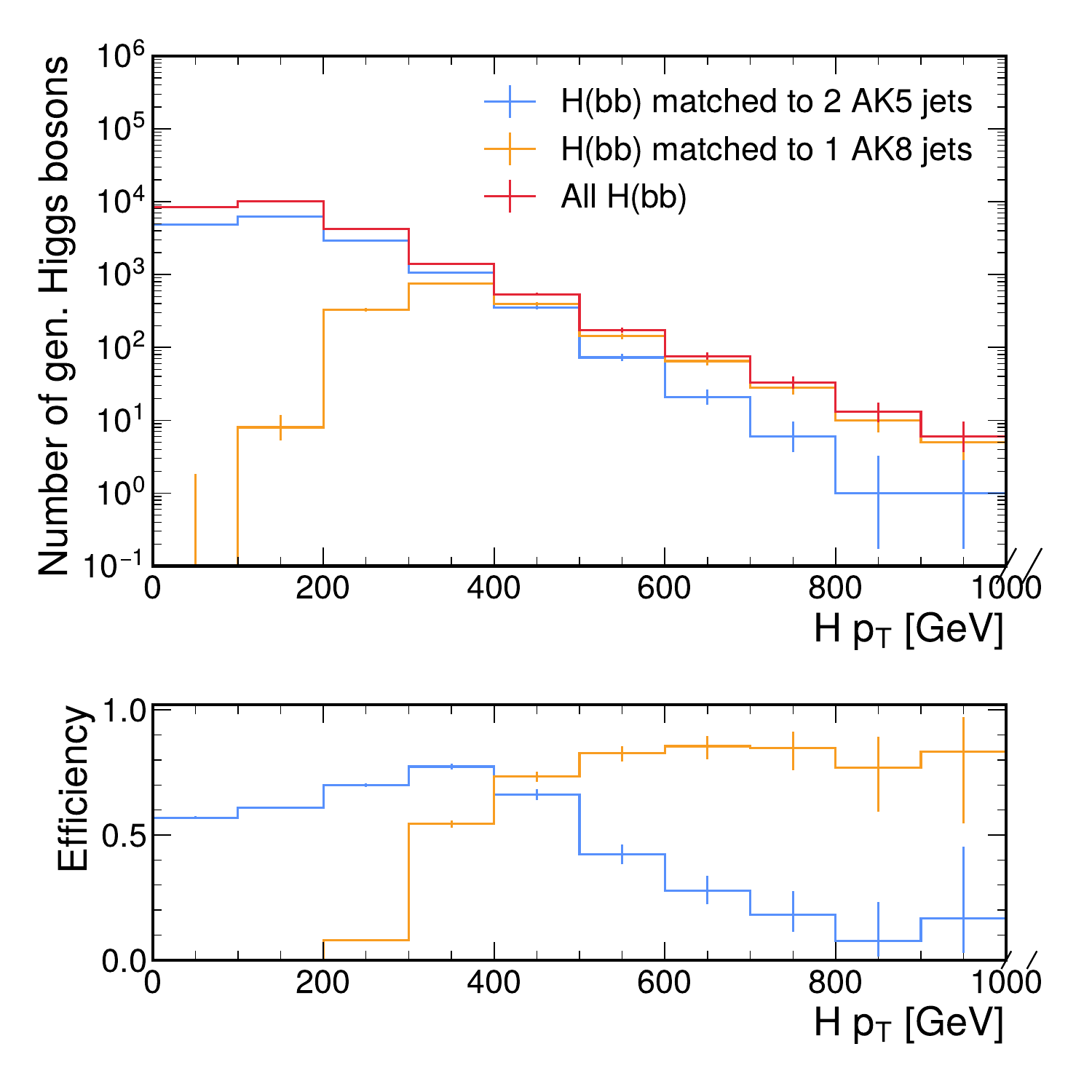}
    \caption{Reconstruction efficiency for true Higgs bosons to be reconstructed and matched to either two AK5 jets or one AK8 jet as a function of \pt in \HHH events with $m_\PH=125\GeV$.}
    \label{fig:higgs_pt}
\end{figure}

The ground truth assignments are obtained by matching the reconstructed jets to the simulated \PQb quarks from the Monte Carlo event record, using a distance measure $\Delta R = \sqrt{\smash[b]{\Delta\phi^{2}+\Delta\eta^{2}}}$.
For AK5 jets, $\Delta R < 0.5$ is required between each jet and \PQb quark daughter, and a pair of jets is labeled as a true Higgs boson if they match to the same Higgs boson in the event record.
For AK8 jets, the jet is labeled a true Higgs boson if $\Delta R < 0.8$ between the jet and the Higgs boson, and between the jet and both \PQb quark daughters.
Multiple Higgs boson candidates are allowed in both reconstructions.
The truth-level matching efficiency of Higgs boson candidates as a function of their \pt is shown in Fig.~\ref{fig:higgs_pt}.
For $\pt > 400\GeV$, AK8 jets have a higher matching efficiency than AK5 jets.
No overlap removal is performed at the training stage.

\HHH events that contain at least six AK5 jets are selected, while \HH events are selected if they contain at least four AK5 jets.
Up to ten AK5 jets and three AK8 jets are considered for both event types.
No $\PQb$-tagging requirements are applied.
Approximately 0.8 million \HHH events and 1.2 million \HH events pass this pre-selection.
These events are split as follows: 90.25\% for training the resolved SPA-Net, 4.75\% for validation and hyper-parameter optimization, and 5\% for testing.

To evaluate any potential distortions in the mass distribution, known as \emph{mass sculpting}, we doubled the SM \HHH dataset and increased the SM \HHH training+validation dataset by simulating additional nonresonant \HHH events with Higgs boson masses set to 120, 122.5, 127.5, and 130\GeV.
Each additional mass point contains the same number of events with the SM \HHH training and validation dataset before pre-selection.
These datasets were combined prior to training the boosted+resolved \HHH SPA-Net models.
This approach aims to produce a flatter Higgs boson mass distribution and reduce mass sculpting effects.

We also generated a background dataset from events composed uniquely of jets produced through the strong interaction, referred to as quantum chromodynamics (QCD) multijet events, using the same generator as for the non-resonant \HH and \HHH events.
To ensure a sufficient sample size with real high-\pt \PQb jets, we generated 4 \PQb partons and required their scalar \pt sum to be $\HT>250\GeV$.
After applying the same pre-selection criteria as for the \HHH events, the QCD test dataset consists of approximately 25,000 events.

A summary of the dataset used in this study is shown Table~\ref{table:dataset}.

\begin{table}
    \resizebox{\linewidth}{!}{
        \begin{tabular}{ccccc}
            \hline
            Sample                                              & $N_{\mathrm{events}}$ & Training Fraction & Validation Fraction & Testing Fraction \\ \hline
            \HHH, $m_\PH=125\GeV$                               & 855,254               & 90.25\%           & 4.75\%              & 5\%              \\
            \HHH, $m_\PH\in\{120, 122.5, 125, 127.5, 130\}\GeV$ & 9,580,718             & 95\%              & 5\%                 & ---              \\
            \HH, $m_\PH=125\GeV$                                & 1,209,807             & 90.25\%           & 4.75\%              & 5\%              \\
            QCD multijet                                        & 127,876               & 80\%              & ---                 & 20\%             \\ \hline
        \end{tabular}
    }
    \caption{
        The number of events $N_{\mathrm{events}}$ after pre-selection.
        Since this study focuses exclusively on SPA-Net's performance on SM signals and backgrounds, we evaluated SPA-Net only on the SM testing dataset ($m_\PH = 125$ GeV). Consequently, the testing fraction of the \HHH dataset, including non-SM mass points, is not applicable. The QCD dataset was used to train the BDT baseline introduced in Section~\ref{sec:baseline} and test the mass sculpting effect of the SPA-Net models and the baselines.
        No validation fraction was allocated for the QCD dataset, as we did not perform hyperparameter tuning for the BDT.}
    \label{table:dataset}
\end{table}

\section{Methods}
\label{sec:methods}

\subsection{Baseline}
\label{sec:baseline}

The resolved training of SPA-Net is benchmarked against the $\chi^2$ method~\cite{Snyder:1995hg}, which minimizes the following quantity:
\begin{equation}
    \chi^2 = \left(m(j_a,j_b)- m_\PH\right)^2 + \left(m(j_c,j_d)- m_\PH\right)^2 + \left(m(j_e,j_f)- m_\PH\right)^2,
\end{equation}
where $m(j_a,j_b)$ is the invariant mass of a pair of AK5 jets $j_a$ and $j_b$ and $m_\PH=125\GeV$ is the nominal mass of the Higgs boson.
The AK5 jets are sorted in descending order by \pt, with the \PQb-tagged sorted jets coming first and the non-\PQb-tagged coming last.
For the $\chi^2$ method, only the first six jets are considered.

\begin{figure}[htpb]
    \centering
    \includegraphics[width=0.5\textwidth]{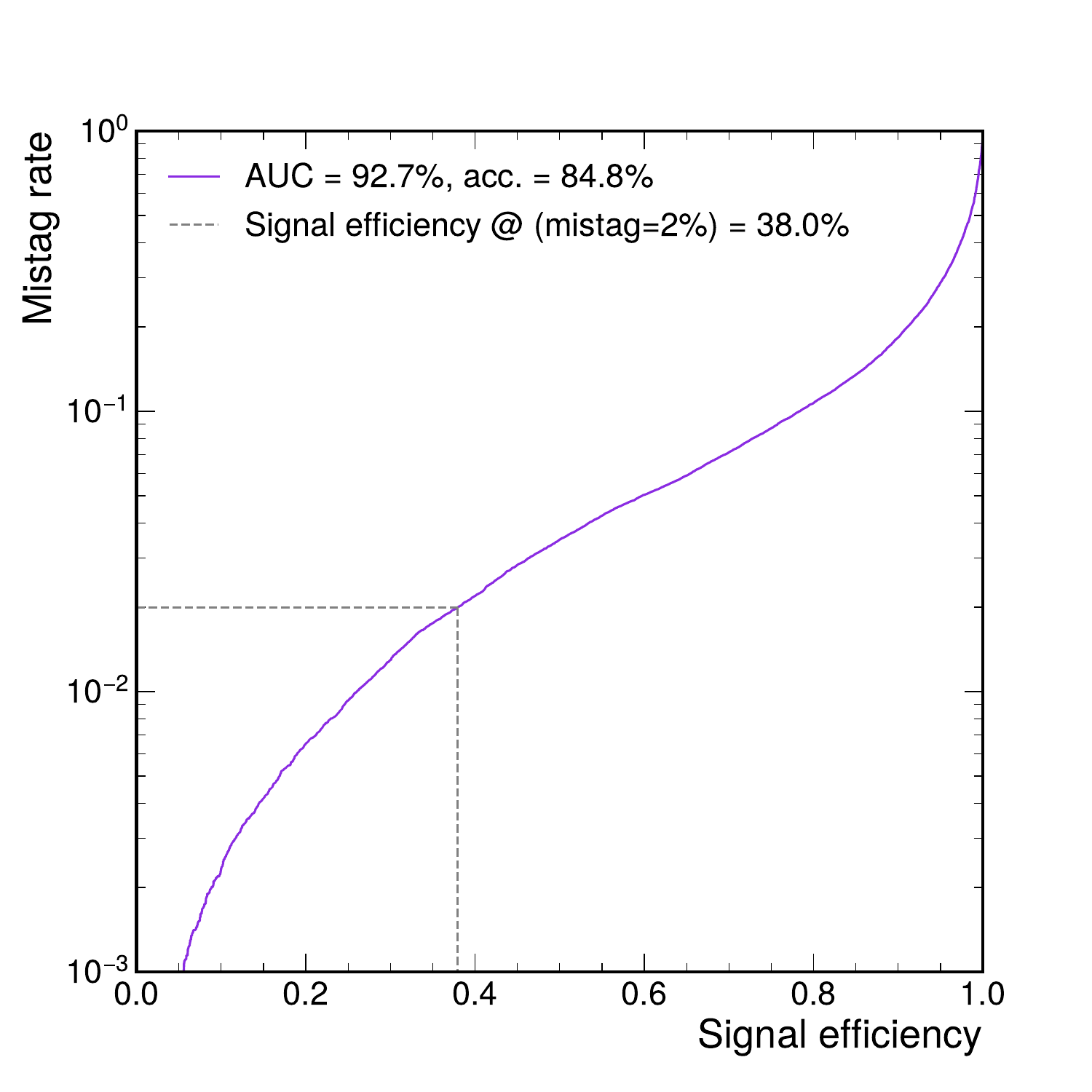}
    \caption{ROC curve for the BDT-based boosted topology baseline, showing the QCD background mistag rate versus the \Hbb signal efficiency.
    The working point corresponds to a signal efficiency of 38\% and a background misidentification rate of 2\%
    }
    \label{fig:bdt_roc}
\end{figure}

For the boosted topology baseline, the default \DELPHES configuration lacks a parameterized boosted Higgs boson AK8 jet tagger and does not save the jet constituents for all AK8 jets necessary to train a state-of-the-art jet tagger like ParticleNet~\cite{Qu:2019gqs} or Particle Transformer~\cite{Qu:2022}.
In lieu of that, we train a boosted decision tree (BDT) tagger as a baseline to compare with SPA-Net.
Given an AK8 jet, the BDT input variables are the \pt, $\eta$, mass $m$, soft-drop mass $m_{\SD}$~\cite{Larkoski:2014wba}, fraction of charged energy, number of charged particles, and the $N$-subjettiness jet substructure variables $\tau_{32}$ and $\tau_{21}$~\cite{Thaler:2010tr}.
The BDT output score represents the probability that the AK8 jet is an \Hbb jet rather than a QCD jet.
It was trained on a dataset that combined the AK8 jets in the SM \HHH training set and the AK8 jets in the QCD training set, which consists of 164,601 \Hbb jets and 553,548 QCD jets.
The classes were balanced during training.
We show the receiver operating characteristic (ROC) curve, qrea under the curve (AUC), and accuracy of the BDT in Fig.~\ref{fig:bdt_roc}.
\subsection{SPA-Net}

SPA-Net~\cite{Shmakov:2021qdz,Fenton:2020woz,Fenton:2023ikr} is a general attention-based neural network architecture for the assignment of reconstructed physics objects (e.g., jets or leptons) to truth-level particles (e.g., Higgs bosons or top quarks).
Since its introduction, there have been several improvements that we leverage in this paper.

\begin{figure}[htpb]
    \centering
    \includegraphics[width=\textwidth]{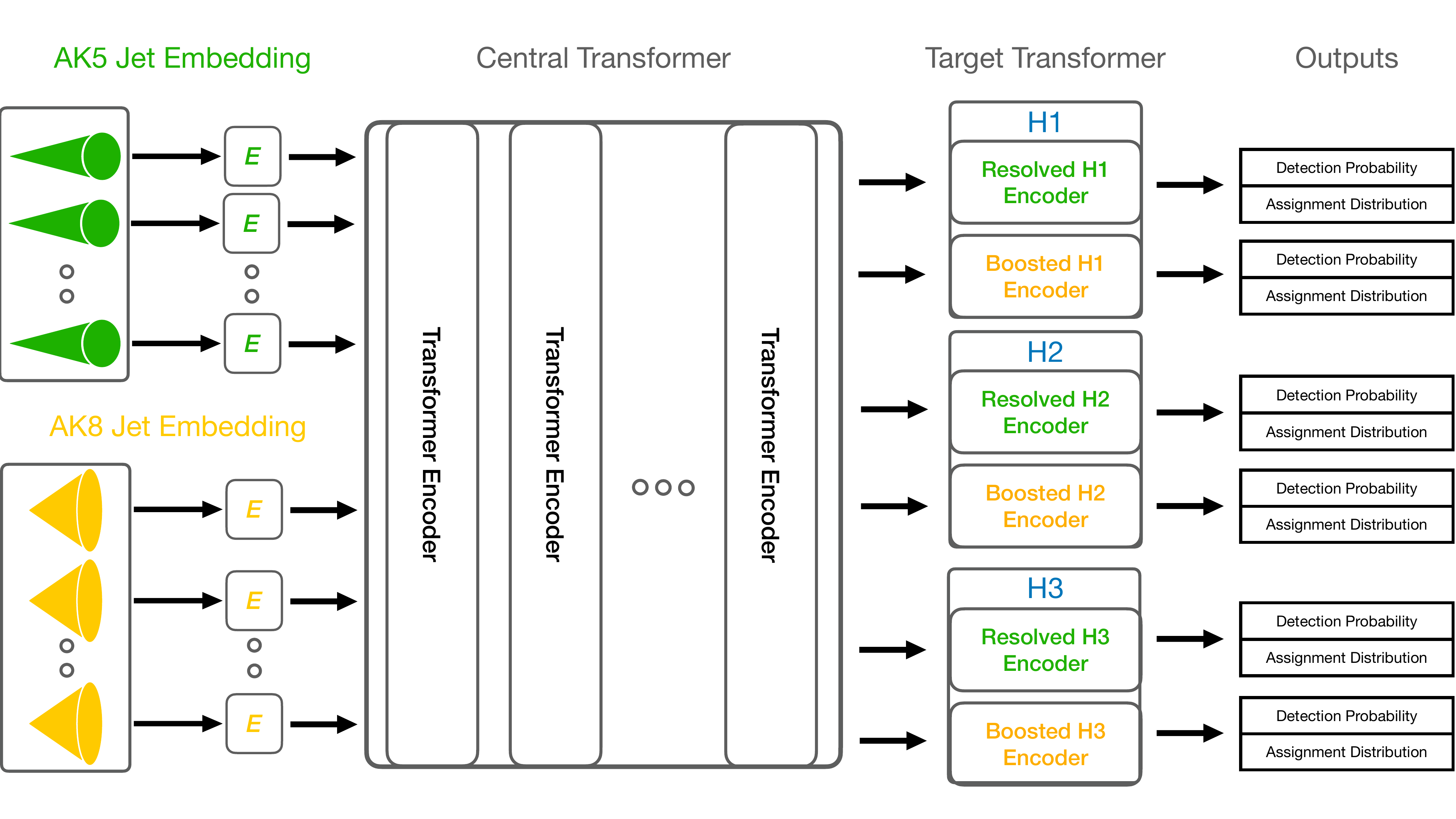}
    \caption{Diagram of the SPA-Net architecture used in this work for \HHH events.
        Both large-radius (AK8) and small-radius (AK5) jets are input to the model with separate  Higgs boson targets.
        The resolved Higgs boson targets consist of pairs of AK5 jets, while the boosted Higgs boson targets consist of single AK8 jets.}
    \label{fig:spanet}
\end{figure}

The structure of SPA-Net, shown in Fig.~\ref{fig:spanet}, consists of five distinct components:
\begin{enumerate}
    \itemsep-0.3em
    \item[(1)] independent object embeddings to produce latent space representations for each reconstructed object;
    \item[(2)] a central stack of transformer encoders;
    \item[(3)] additional transformer encoders for each target;
    \item[(4)] a tensor-attention to produce the object-target assignment distributions; and
    \item[(5)] a classification head for target detection.
\end{enumerate}
The transformer encoders employ multi-head self-attention~\cite{transformers} with one significant modification: the positional embeddings are combined with \textit{position-independent} physics object embeddings which preserve permutation invariance in the input.

While the original SPA-Net studies concentrated on examples where all physics objects were of the same type (e.g., small-radius jets), it was updated in Ref.~\cite{Fenton:2023ikr} to allow for the consideration of different types of physics objects, specifically leptons.
In this application, we leverage this improvement with the novelty of considering both small-radius and large-radius jets, which may both be associated to the same truth-particle.

We accommodate these additional inputs by training individual position-independent embeddings for each \textit{class} of input.
This allows the network to adjust to the various distributions for each input type, and allows us to define sets of features specific to  each type of object.
The individual embedding layers map these disparate objects with different features into a unified latent space, which may be processed by the central transformer.
The encoded event vector after the central transformer is a latent summary representation of the entire event.



Depending on its transverse momentum, a Higgs boson can be reconstructed in a resolved topology, meaning two AK5 jets, or  a boosted topology, meaning a single AK8 jet.
Moreover, these two possibilities are not mutually exclusive in that a single Higgs boson may be reconstructible in both ways.
Motivated by the two different topologies, we replaced the particle transformer branches in the original SPA-Net~\cite{Shmakov:2021qdz,Fenton:2020woz} by target transformer branches to enable the same particle to be reconstructed in two different classes of objects.
To account for the two possible topologies for each particle, we designed a resolved target and a boosted target for each potential Higgs boson in an event.
Each target's branch follows the same architecture as the ``particle'' branch illustrated in Ref.~\cite{Shmakov:2021qdz} and includes a tensor attention output and a binary detection output.

The tensor attention output is interpreted as the assignment probability (\AP).
Each entry of the tensor represents the likelihood that a set of objects reconstructs the target.
The indices of the tensor entry indicate which objects are to be assigned to the target.
To train SPA-Net to output the assignment tensor, we adopt the combined symmetric loss from the original SPA-Net~\cite{Shmakov:2021qdz} and disable the class-balance term, as we found it degraded performance on the majority of events.
Explicitly, the assignment loss is
\begin{equation}
    \label{equ:ass_loss}
    \mathcal{L}_
    \mathrm{assignment} = \min_{\sigma \in G_t}\sum_{i=1}^{N_\mathrm{target}}\mathcal{M}_{\sigma(i)}\mathrm{CCE}(P_i,T_{\sigma(i)}),
\end{equation}
where $\mathrm{CCE}$ represents the categorical cross-entropy loss, $G_t$ denotes the permutation group of all targets, restricted to permutations within the same reconstruction topology, $\sigma$ is an element of the permutation group, $N_\mathrm{target}$ represents the maximum number of the targets that the SPA-Net model can output, $P_i$ is the $i$th predicted jet assignment, $T_{\sigma(i)}$ is the $i$th target jet assignment under permutation $\sigma$, and $\mathcal{M}_{\sigma(i)}$ is the target mask associated with $T_{\sigma(i)}$.
The assignment outputs are trained only on examples in which the event contains all detector objects necessary for a correct target assignment, i.e. the target is  \textit{reconstructible}.
Nonreconstructible targets are ignored via the mask $\mathcal{M}_{\sigma(i)}$ in Eq.~(\ref{equ:ass_loss}).
As a result, the SPA-Net \emph{assignment probability} only represents a conditional assignment distribution over jet indices for each target given that the target is reconstructible.

Since Ref.~\cite{Fenton:2020woz} introduced the ability to reconstruct partial events, it is important to estimate the probability that a given target is reconstructible in the event.
This \emph{detection probability (DP)} is estimated with an additional output head of SPA-Net.
We also take into account the event-level symmetries in a similar manner to the assignment loss.
Specifically, a summary \textit{target vector} is extracted from each of the target transformer encoders.
These target vectors are fed into a feedforward binary classification network to produce a detection probability for each target.
This target detection output is trained with the detection loss,
\begin{equation}
    \label{equ:det_loss}
    \mathcal{L}_
    \mathrm{detection} = \min_{\sigma \in G_t}\sum_{i=1}^{N_\mathrm{target}}\mathrm{BCE}(\DP_i,\mathcal{M}_{\sigma(i)}),
\end{equation}
where BCE denotes the binary cross-entropy loss, and $\DP_i$ is the predicted detection probability of the $i$th target.

The complete loss equation for the entire network is given by
\begin{equation}
    \mathcal{L} =
    \alpha_0 \mathcal{L}_{\mathrm{assignment}} + \alpha_1 \mathcal{L}_{\mathrm{detection}},
\end{equation}
where $\alpha_0$ and $\alpha_1$ are the weights of the different components of the loss function.
In this work, we use $\alpha_0=20$ and $\alpha_1=1$.

\subsection{Target Higgs Boson Mass Reweighting}

\begin{figure}[htpb]
    \centering
    \includegraphics[width=0.3\textwidth]{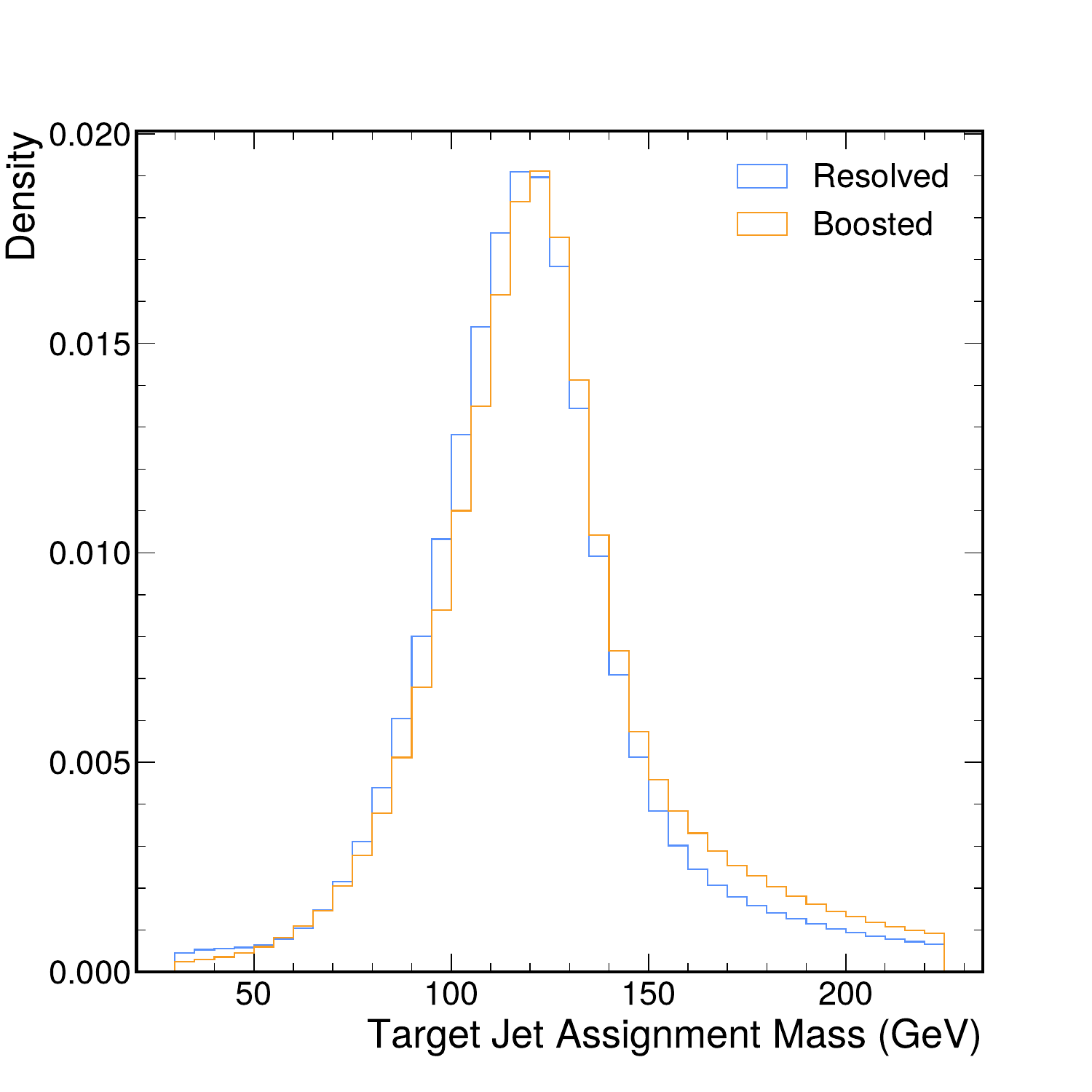}
    \includegraphics[width=0.3\textwidth]{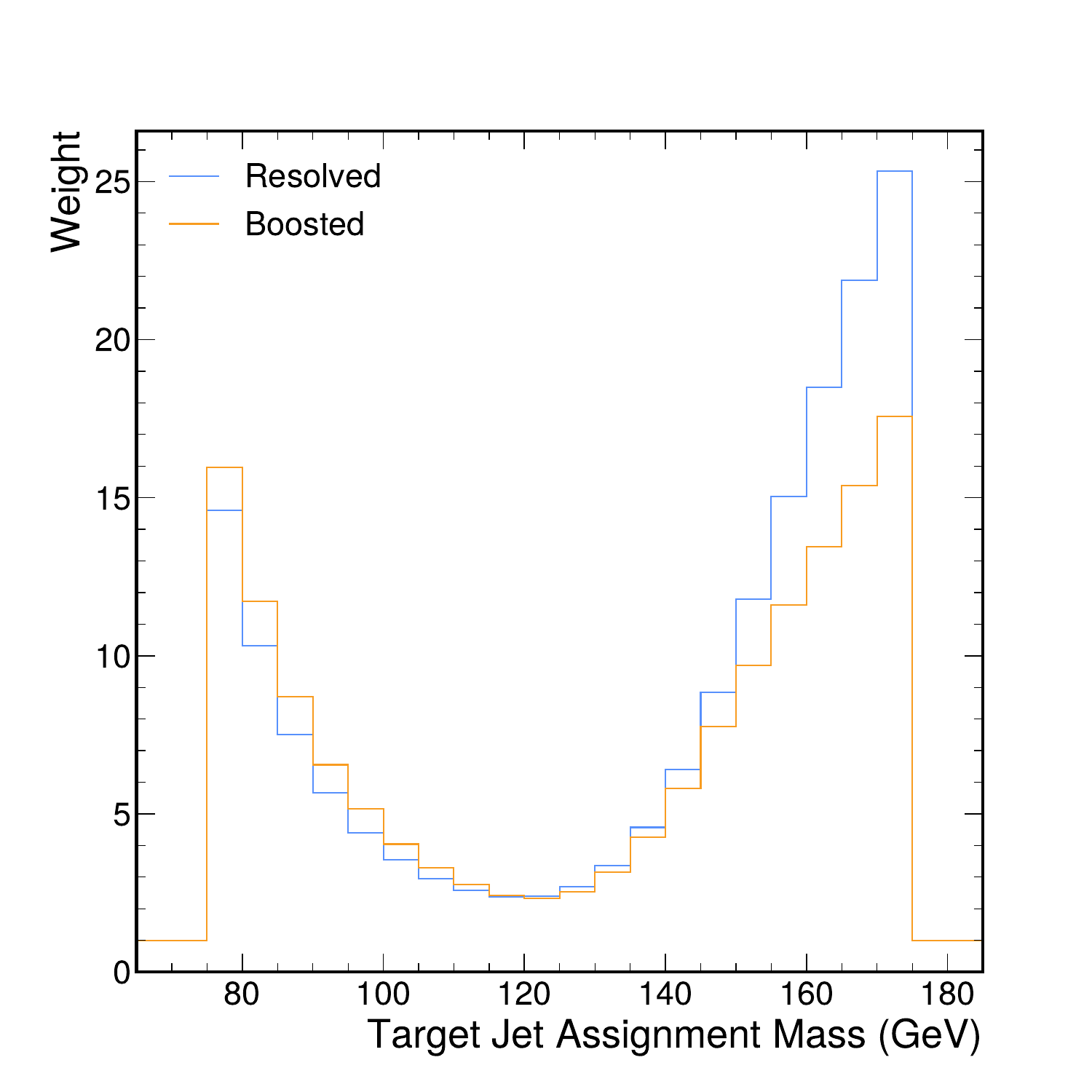}
    \includegraphics[width=0.3\textwidth]{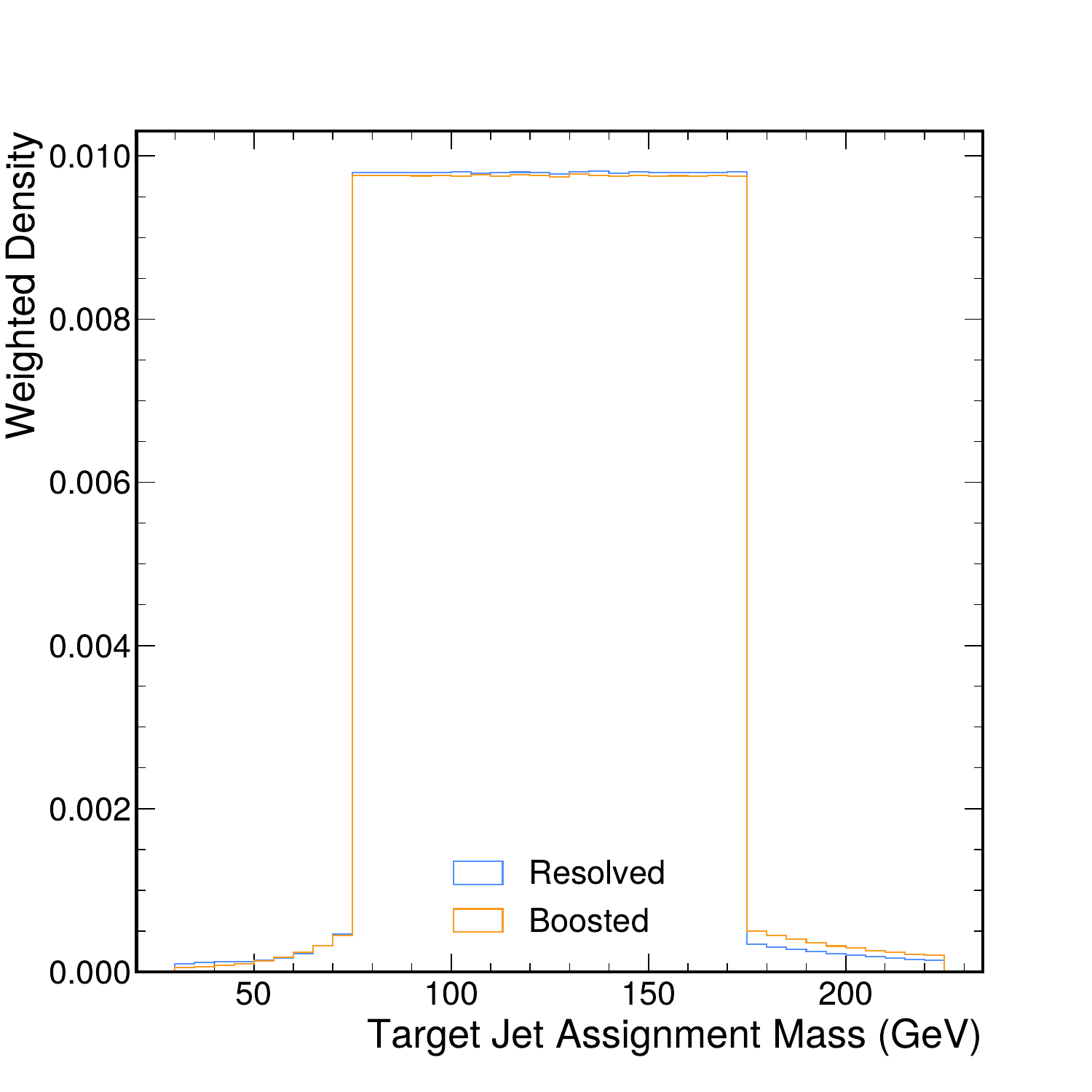}
    \caption{
        The mass distributions of target jet assignments show  peak-like shapes, despite the Higgs targets being generated at uniformly spaced mass points (left).
        The weights derived to flatten the mass spectrum in the range $m\in[75, 175]\GeV$ (center).
        Outside the mass range $[75, 175] \GeV$, the weights are one.
        The resulting mass distribution after reweighting (right).}
    \label{fig:target_mass_dist}
\end{figure}

The mass distribution of the resolved and boosted Higgs boson targets in the combined \HHH dataset is shown in Fig.~\ref{fig:target_mass_dist} (left).
Despite varying the Higgs boson mass between 120 and 130\GeV,
both the resolved and boosted targets still possess an obvious peak around $m_\PH=125\GeV$.
An overabundance of samples at $m_\PH=125\GeV$ in the training set could bias the SPA-Net algorithm to reconstruct Higgs bosons by simply choosing the jet assignment where the invariant mass is closest to $m_\PH=125\GeV$.
This behavior is undesirable because a reconstruction that heavily relies on the invariant mass can result in \emph{mass sculpting} of the background, that is introducing an artificial signal-like peak in the invariant mass distribution even when no signal is present.

To further prevent SPA-Net from reconstructing Higgs bosons only based on the invariant mass of the jet assignment, we implemented a binned reweighting as a function of the Higgs boson target mass $m\in[75, 175]\GeV$.
The weights were calculated independently for each topology and were incorporated in the loss term calculation.
For each topology and each mass bin of width 5\GeV, we calculated the weight as
\begin{equation}
    \label{equ:weight}
    w(m) = \begin{cases} \displaystyle\frac{p_\mathrm{uniform}(m_i)}{p_\mathrm{mass}(m_i)}, & 75 \leq m_{i} \leq m < m_{i+1} \leq175\GeV \\
              1,                                                                 & m \notin [75,175]\GeV
    \end{cases},
\end{equation}
where $p_\mathrm{mass}$ is the real mass distribution and $p_\mathrm{uniform}$ is the desired uniform distribution, and $m_i$ and $m_{i+1}$ are bin edges.
The calculated weight for each bin for each topology is shown in Fig.~\ref{fig:target_mass_dist} (center).
We assigned a weight of one to the targets outside the mass range [75, 175]\GeV to prevent assigning a high weight to outliers.
The weights are incorporated into the calculations for both the assignment loss and the detection loss.
For instance, the assignment loss for each event is modified to be
\begin{equation}
    \label{equ:new_ass_loss}
    \mathcal{L}_
    \mathrm{assignment} = \min_{\sigma \in G_t}\sum_{i=1}^{N_\mathrm{target}}w_{\sigma(i)}\mathcal{M}_{\sigma(i)}\mathrm{CCE}(P_i,T_{\sigma(i)}),
\end{equation}
where, in contrast to Eq.~(\ref{equ:ass_loss}), the assignment loss term for each target is scaled by the corresponding target-wise weight $w_{\sigma(i)}$.
Similarly, the detection loss is modified by multiplying the loss term for each target in the event by $w_{\sigma(i)}$ .

The target weights were also integrated into the validation metrics.
During training, the best checkpoint was selected based on the weighted jet assignment accuracy (JA), averaged across the events in the validation dataset.
JA is defined as
\begin{equation}
    \mathrm{JA} =
    \frac{1}{\sum_{\mathrm{i}=1}^{N_{\mathrm{target}}}w_i}
    \max_{\sigma\in G_t}
    \sum_{i=1}^{N_\mathrm{target}}w_{\sigma(i)}
    \left[P_i\equiv T_{\sigma(i)}\right],
\end{equation}
where $[x\equiv y]$ represents a function that outputs 1 if the jet assignments $x$ and $y$ are equivalent, and 0 otherwise.
Another key difference between our validation metric and the default SPA-Net validation metric lies in the selection of events used for the computation.
The default SPA-Net validation accuracy is averaged only across events in which all possible target particles are reconstructible (fully reconstructible events).
However, we found that we needed to include partially reconstructible  events as well when incorporating boosted and resolved targets because only a small fraction of events contain three Higgs bosons that can be reconstructed in both topologies.


\section{Postprocessing}
\label{sec:postprocessing}


We postprocess the outputs for the two topologies separately.
For each topology, we calculate the most probable number of Higgs bosons in that topology based on the multinomial distribution.
The probabilities for detecting a specific number of Higgs bosons, $N_\PH$, are calculated as:

\begin{align}
    p(N_\PH) & = \begin{cases}
    (1-\DP_1) (1-\DP_2) (1-\DP_3) & N_\PH = 0 \\
     \DP_1(1-\DP_2)(1-\DP_3) +\DP_2 (1-\DP_1)(1-\DP_3)  & N_\PH = 1 \\
               {}+ \DP_3 (1-\DP_1)(1-\DP_2)\\
   \DP_1 \DP_2(1-\DP_3)+\DP_2 \DP_3 (1-\DP_1)  +\DP_3 \DP_1(1-\DP_2) & N_\PH = 2 \\
    \DP_1 \DP_2  \DP_3 & N_\PH=3
    \end{cases}.
\end{align}


Then, we obtain the most probable number of Higgs bosons in the topology, $\hat N_\PH$, as
\begin{equation}
    \label{equ:nh}
    \hat N_\PH = \argmax_{N_\PH}p(N_\PH),
\end{equation}
Subsequently, we select the top $\hat N_\PH$ Higgs boson's jet assignments ranked by the product $\DP_i \AP_i$, which combines the detection probability (whether the Higgs boson exists in this topology) with the assignment probability (the confidence of the jet assignment) to prioritize the most reliable candidates.

Additionally, to accurately analyze Higgs bosons originating from both resolved and boosted topologies, it is crucial to implement strategies that prevent double counting.
Without such measures, the same Higgs boson could be reconstructed in both topologies and appear twice in different categories.
In this study, we address this issue by prioritizing boosted Higgs boson candidates, as these typically have a significantly better signal-to-background ratio~\cite{Butterworth:2008iy,CMS:2017bcq}.
Specifically, resolved Higgs boson candidates are discarded if any of their associated AK5 jets overlap with an AK8 jet corresponding to a boosted Higgs boson candidate.
The overlap condition is defined using a separation criterion of $\Delta R < 0.5$.
Similarly, for the purpose of uniquely identifying a (boosted or resolved) Higgs boson target for post-training evaluation metrics, resolved Higgs boson targets are discarded if the same Higgs boson also matches an AK8 jet according to the criteria in Section~\ref{sec:dataset}.





\section{Results}
\label{sec:results}


The algorithm's performance is quantified using two metrics: \textit{reconstruction efficiency} and \textit{reconstruction purity}.
Reconstruction efficiency refers to the fraction of target Higgs bosons that are correctly reconstructed by the SPA-Net jet assignment.
Conversely, reconstruction purity is defined as the fraction of reconstructed Higgs bosons that match the target Higgs bosons.
Here, the matching condition requires that the candidate and target Higgs bosons have the same jet assignment.
These metrics are computed per particle target, rather than at the event level.

In this section, we present results for the following SPA-Net trainings:
\begin{enumerate}
    \itemsep-0.3em
    \item \HHH events targeting (a) a resolved-only topology considering up to 10 AK5 jets and (b) resolved and boosted topologies, the latter considering up to 3 AK8 jets.
    \item \HH events targeting (a) a resolved-only topology considering up to 10 AK5 jets and (b) resolved and boosted topologies, the latter considering up to 2 AK8 jets.
\end{enumerate}
The training configurations are shown in Table~\ref{table:train_config}.

\begin{table}[htpb]
    \caption{The training configurations of the models presented in Section~\ref{sec:results} are described.
    }
    \resizebox{\linewidth}{!}{
        \begin{tabular}{l|cccc}
            \multirow{2}{*}{\textbf{Parameter}} & \multicolumn{4}{c}{\textbf{Benchmark Models}}                                                       \\
                                                & \HHH Boosted+Resolved                         & \HHH Resolved & \HH Boosted+Resolved & \HH Resolved \\ \hline
            Training Epochs                     & 500                                           & 500           & 500                 & 500         \\
            Learning Rate                       & 0.0005                                        & 0.0005        & 0.0015               & 0.0015       \\
            Batch Size                          & 4096                                          & 4096          & 4096                 & 4096         \\
            Dropout                             & 0.2                                           & 0.2           & 0.2                  & 0.2          \\
            L2 Gradient Clipping                & 1.0                                           & 1.0           & 1.0                  & 1.0          \\
            L2 penalty                          & 0.0002                                        & 0.0002        & 0.0002               & 0.0002       \\ \hline
            Hidden Dimension                    & 64                                            & 32            & 32                   & 32           \\
            Central Encoders                    & 8                                             & 8             & 8                    & 8            \\
            Branch Encoders                     & 6                                             & 6             & 6                    & 6            \\
            Number of heads & 8 & 8 & 8 & 8 \\ \hline
            Partial Event Training              & Yes                                           & Yes           & Yes                  & Yes          \\
            Cosine Annealing Cycles             & 5                                             & 1             & 1                    & 1            \\
        \end{tabular}
    }
    \label{table:train_config}
\end{table}

\subsection{\texorpdfstring{\HHH}{HHH} Resolved Training}

\begin{figure}[htpb]
    \centering
    \includegraphics[width=\textwidth]{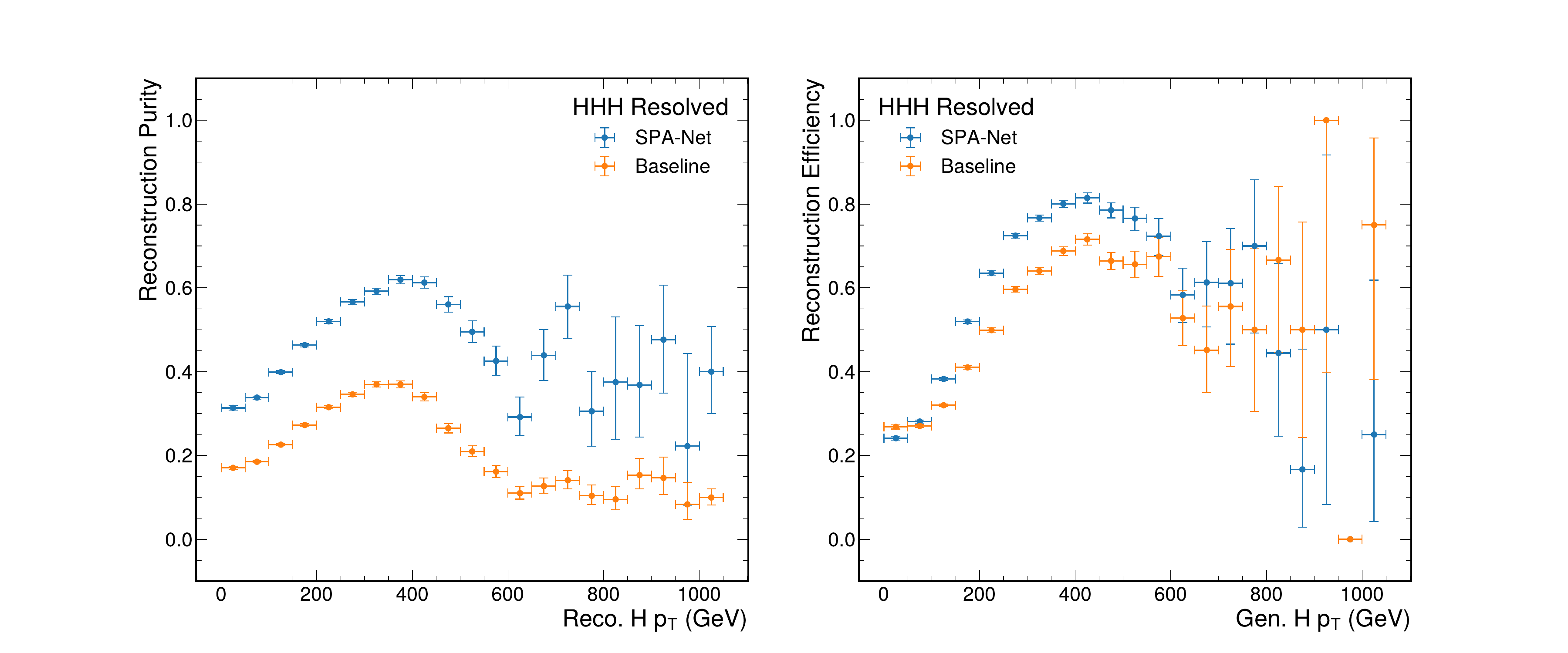}
    \caption{On the left, Reco.~\PH refers to the Higgs boson candidates reconstructed by the SPA-Net jet assignments.
        The reconstruction purity is defined as the fraction of the Higgs boson candidates that have the same jet assignment as the target.
        On the right, Gen.~\PH refers to the Higgs boson targets consisting of the target jet assignments.
        The reconstruction efficiency is defined as the fraction of the target Higgs bosons recovered by the SPA-Net predictions.
        The error bar in each bin is the Clopper-Pearson interval.}
    \label{fig:higgs_hhhr}
\end{figure}

Figure~\ref{fig:higgs_hhhr} presents the Higgs reconstruction purity and efficiency as functions of the Higgs boson candidate transverse momentum \pt for the fully resolved SPA-Net training.
Below $\pt<600\GeV$, SPA-Net demonstrates a higher reconstruction efficiency than the $\chi^2$ baseline, except in the first bin.

As described in Section~\ref{sec:postprocessing}, we discard jet assignments where $\hat N_\PH<3$, which indicates that not all Higgs bosons could be reconstructed.
This procedure aims to balance reconstruction purity and efficiency and results in enhanced reconstruction purity with respect to the baseline, as shown in Fig.~\ref{fig:higgs_hhhr}.
However, this balancing strategy can occasionally discard valid jet assignments, reducing the reconstruction efficiency.

The jet assignment performance is affected by the angular separation of the two jets from the Higgs boson decay, which depends on the Higgs boson momentum.
For low \pt, the jets are widely separated and distributed across the detector, making the jet assignment challenging for three Higgs bosons.
For high \pt, the jets are more collimated and the jet correlation is more evident.
However, above $\pt > 400\GeV$, the efficiency drops due to the increased probability of merging two jets into a large-radius jet or losing one of the \PQb jets out of the detector acceptance.
This limitation can be overcome by incorporating the large-radius AK8 jets in the SPA-Net training.

\subsection{\texorpdfstring{\HHH}{HHH} Boosted+Resolved Training}

\begin{figure}[htpb]
    \centering
    \includegraphics[width=1\textwidth]{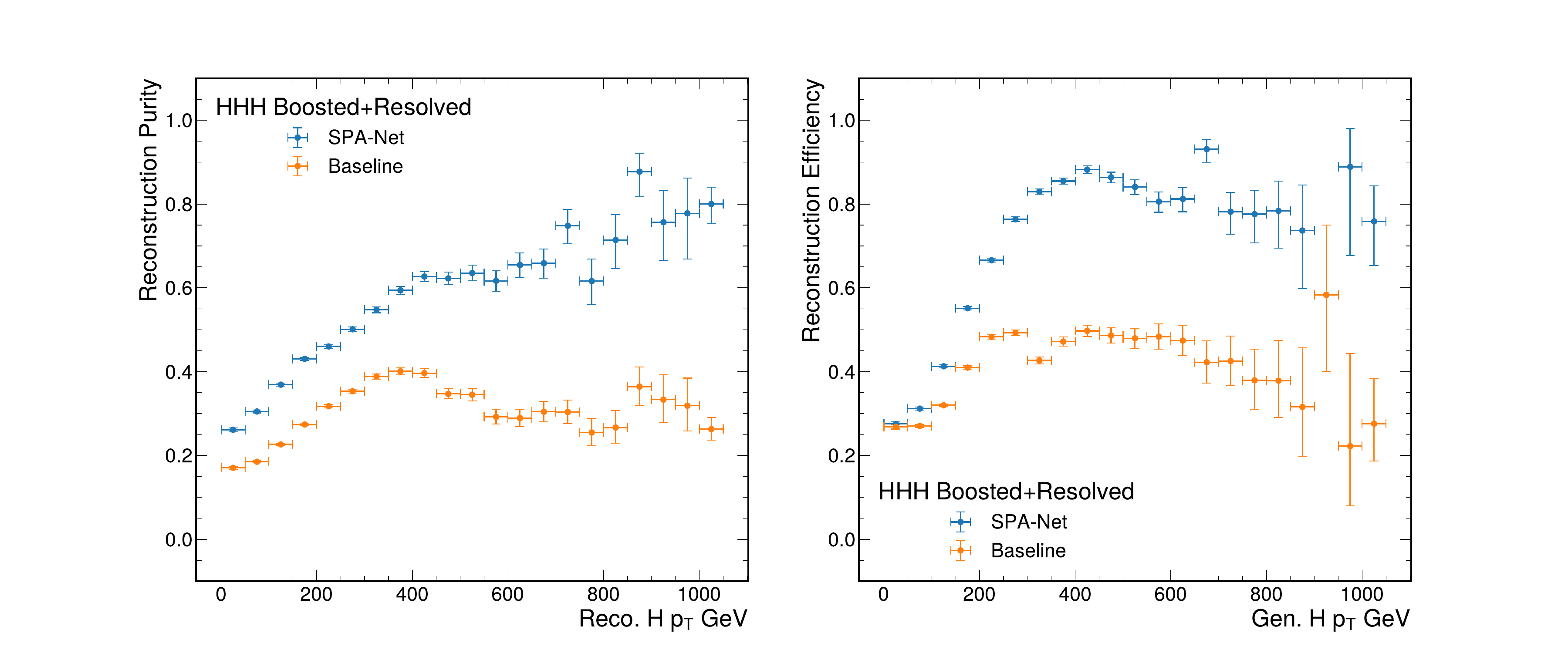}
    \caption{SPA-Net's performance improves in the high-\pt region when boosted targets are incorporated during training.
        A comparison of boosted+resolved SPA-Net with the $\chi^2$+BDT baseline shows that SPA-Net consistently outperforms the baseline across both evaluation metrics.
        The axis definitions are the same as those in Fig.~\ref{fig:higgs_hhhr}.}
    \label{fig:higgs_hhhbr}
\end{figure}

Figure~\ref{fig:higgs_hhhbr} shows the Higgs reconstruction purity and efficiency versus the Higgs boson candidate transverse momentum \pt for the SPA-Net model trained on both resolved and boosted targets.
Compared with Fig.~\ref{fig:higgs_hhhr}, the boosted+resolved SPA-Net exhibits significantly better performance above $\pt > 400\GeV$ as a result of including boosted targets in the training.

As introduced in Section~\ref{sec:baseline}, the baseline utilizes the $\chi^2$ method to assign AK5 jets to resolved Higgs boson targets and applies the BDT to classify AK8 jets into boosted Higgs boson candidates or background.
Similar to SPA-Net, which  predicts and reconstructs candidates in both topologies for each event, the $\chi^2$-based resolved baseline and the BDT-based boosted baseline also operate on each event.
In this study, we assume the $\chi^2$ baseline always predicted three resolved Higgs boson candidates in the resolved topology.
For the boosted topology, we select the AK8 jets of which BDT scores passing the loose working point of 0.911348, corresponding to a signal efficiency of 38\% and a background misidentification rate of 2\%, as shown in Fig.~\ref{fig:bdt_roc}.
Finally, the Higgs boson candidates predicted by both baselines undergo the same overlap-removal procedure in Section~\ref{sec:postprocessing}.

Across all Higgs boson $\pt$ bins, the boosted+resolved SPA-Net exhibits higher Higgs boson reconstruction purity and efficiency than the $\chi^2$ and BDT baseline.

\begin{figure}[htpb]
    \centering
    \includegraphics[width=0.45\textwidth]{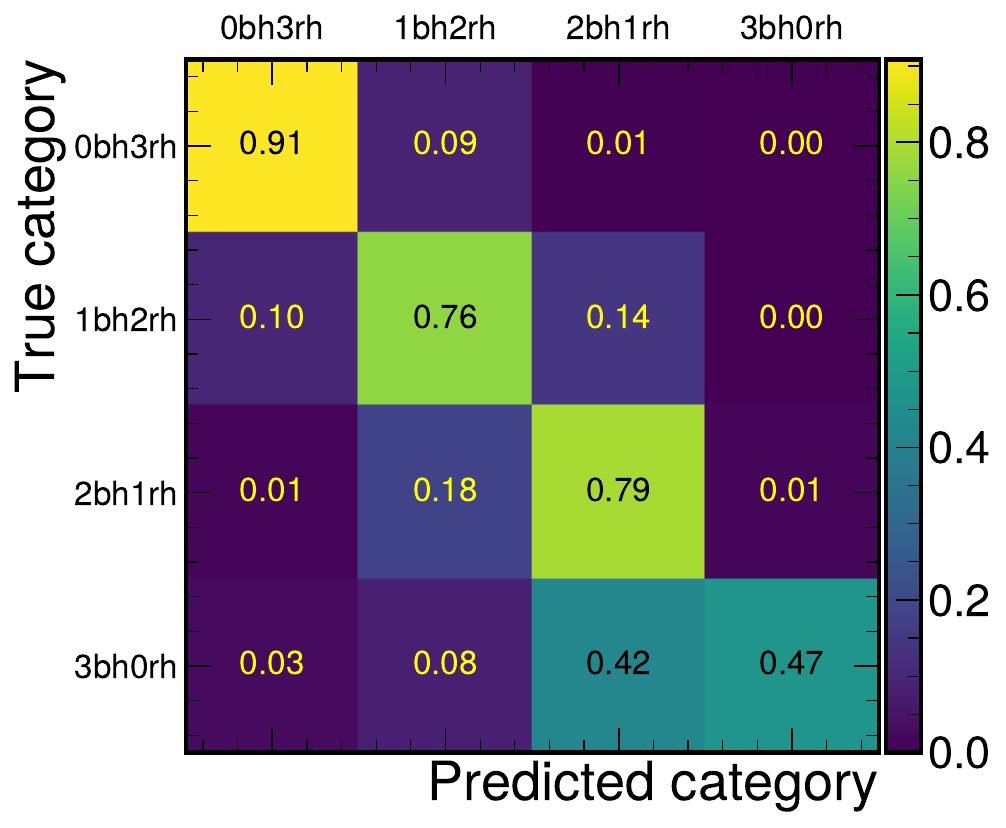}
    \caption{Confusion matrix for the full events in the testing fraction of the \HHH $m_\PH = 125\GeV$ dataset.
    The true label indicates the number of boosted and resolved-only targets in each event, while the predicted label corresponds to the number of boosted and resolved Higgs boson candidates identified after postprocessing.
    Most true categories maximized along the diagonal, indicating strong event topology classification performance.}
    \label{fig:hhh_confusion_matrix}
\end{figure}

Additionally, we present the confusion matrix for events with three reconstructible Higgs bosons target (referred to as full events) in Fig.~\ref{fig:hhh_confusion_matrix}.
An event is considered full if all true Higgs bosons can be successfully matched to either 2 AK5 jets or 1 AK8 jet as described in Section~\ref{sec:dataset}.
We focus on full events as they are expected to dominate the signal sensitivity in experimental searches.
The classification categories are as follows: three resolved target Higgs bosons (0bh3rh), one boosted and two resolved targets (1bh2rh), two boosted and one resolved target (2bh1rh), and three boosted targets (3bh0rh).
For each event, the true category is determined by counting the number of boosted targets and the number of resolved targets that do not correspond to any boosted target.
The predicted category is obtained by counting the number of boosted candidates and resolved candidates after applying the postprocessing procedure described in Section~\ref{sec:postprocessing}.
The matrix is normalized over the true category (rows).

Most true categories are predicted correctly, indicating the strong event topology classification performance of SPA-Net and our postprocessing method.
An exception is the 3bh0rh category, which is underrepresented in the dataset (0.1\%), leading to reduced prediction accuracy.

\subsubsection{Mass Sculpting}

\begin{figure}[htpb]
    \centering
    \includegraphics[width=0.5\textwidth]{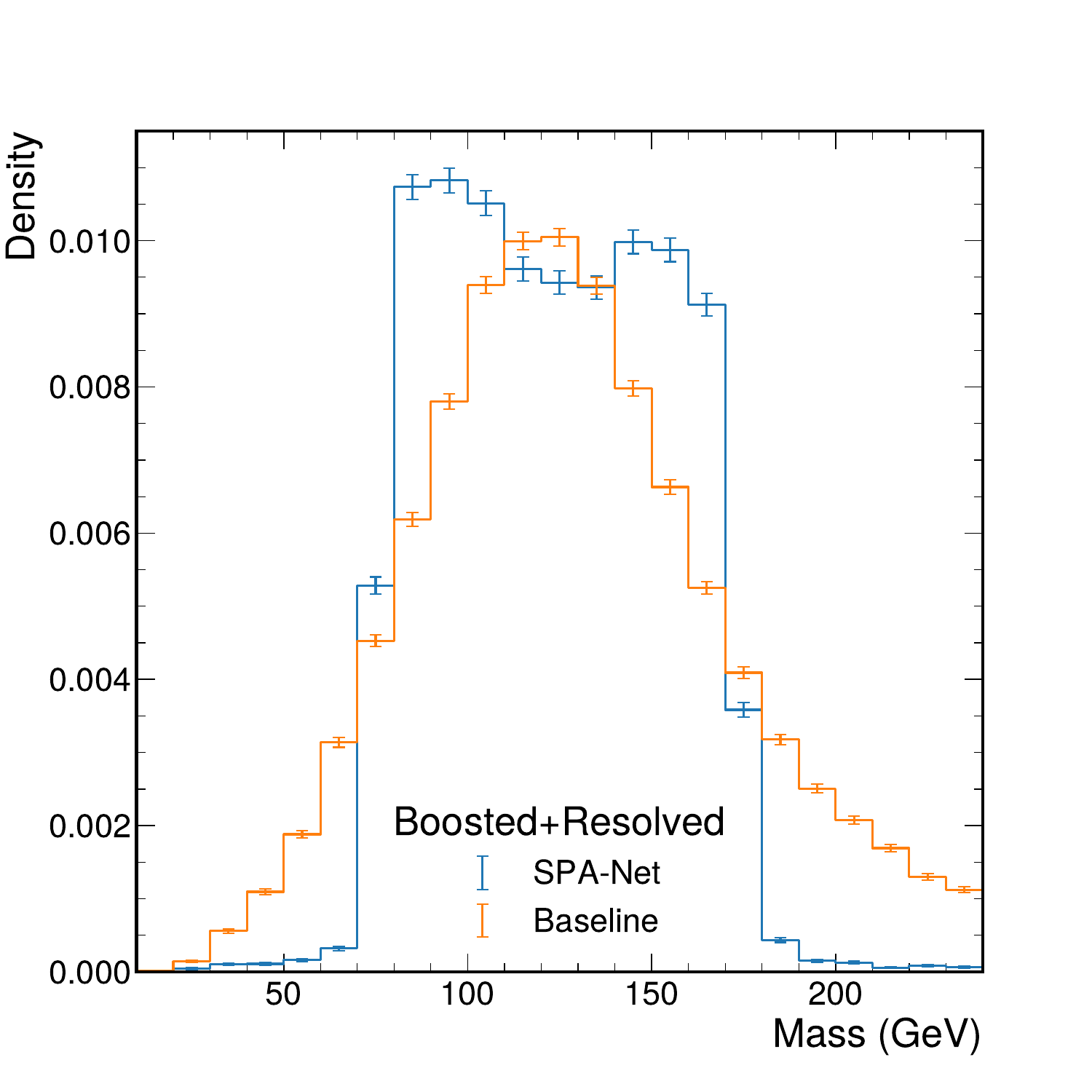}
    \caption{The boosted+resolved SPA-Net model, trained with the calculated weights, outperforms the $\chi^2$+BDT baseline in reducing the correlation between the jet assignments and the reconstructed Higgs boson candidate mass.}
    \label{fig:mass_sculpting}
\end{figure}

As specified at the end of Section~\ref{sec:dataset}, the training dataset of the boosted+resolved SPA-Net includes Higgs bosons simulated at mass points different than 125\GeV.
We ran inference of the boosted+resolved SPA-Net model on the QCD test dataset introduced in Section~\ref{sec:dataset}. Figure~\ref{fig:mass_sculpting} shows the mass distribution of the Higgs boson candidates predicted by the boosted+resolved SPA-Net model is smoother and does not peak at 125\GeV compared to that of the $\chi^2$+BDT baseline, indicating the boosted+resolved SPA-Net model distorts the mass distribution less than the baseline.

\subsection{\texorpdfstring{\HH}{HH} Resolved and Boosted+Resolved Trainings}

\begin{figure}[htpb]
    \centering
    \includegraphics[width=1\textwidth]{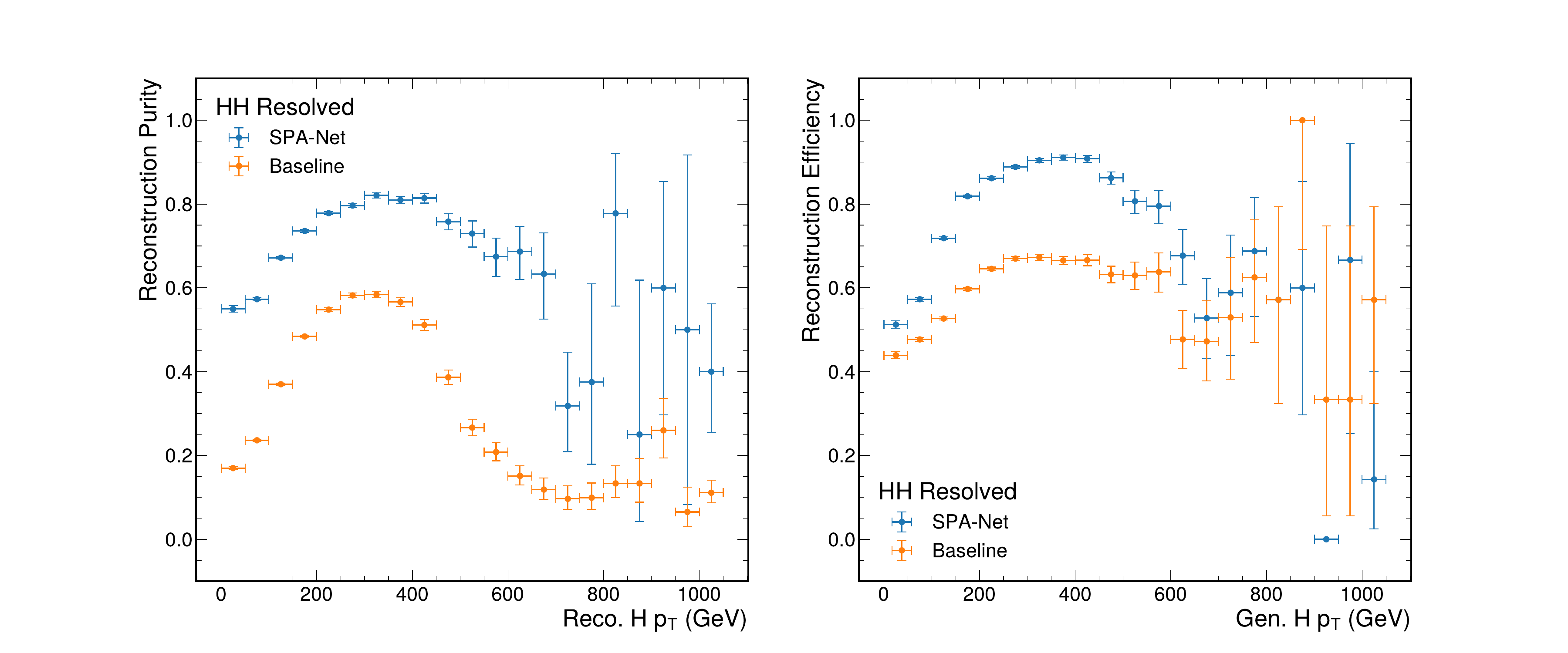}
    \caption{Our resolved SPA-Net trained on \HH targets consistently outperforms the $\chi^2$ baseline across both evaluation metrics.
        The axis definitions are the same as those in Fig.~\ref{fig:higgs_hhhr}.}
    \label{fig:higgs_hhr}
\end{figure}

\begin{figure}[htpb]
    \centering
    \includegraphics[width=1\textwidth]{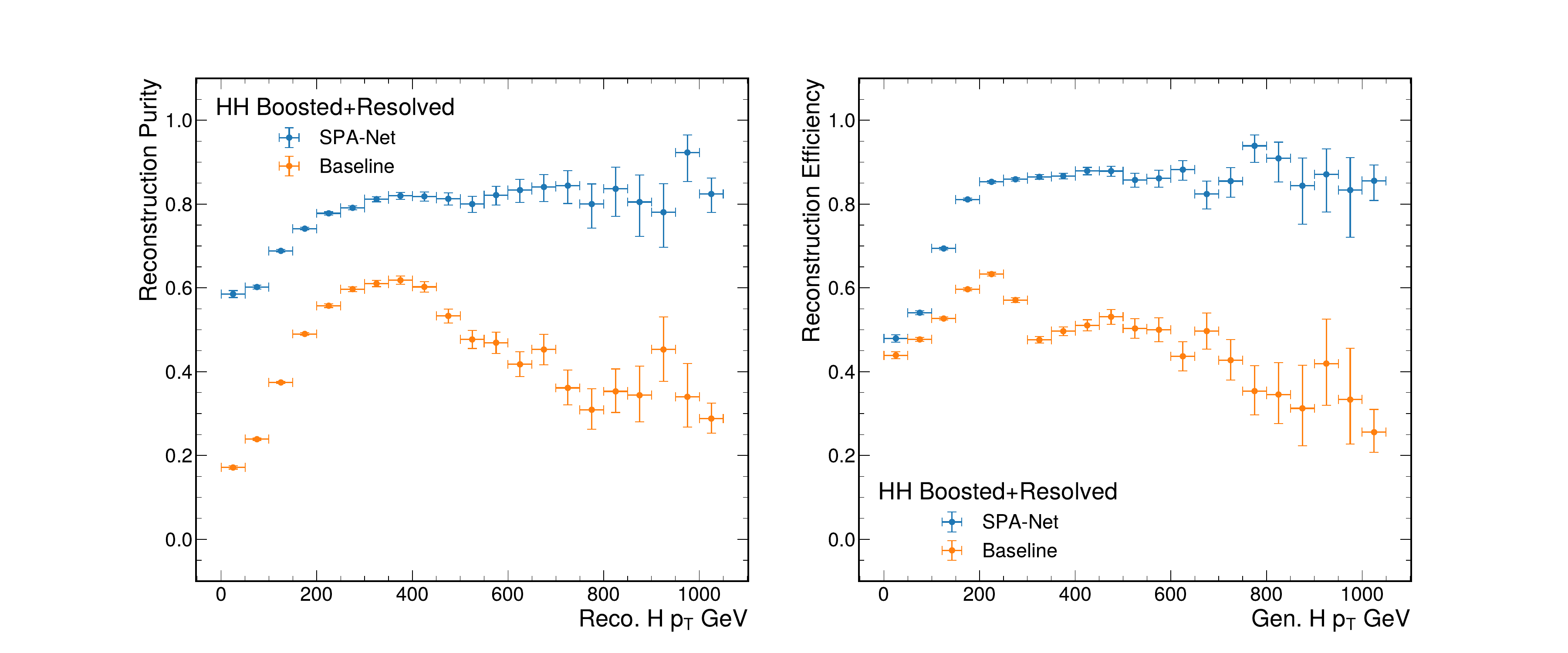}
    \caption{Similarly, compared to Figure~\ref{fig:higgs_hhr}, SPA-Net's performance improves in the high-\pt region when boosted targets are incorporated during training.
        A comparison of boosted+resolved SPA-Net training with the $\chi^2$+BDT baseline shows that SPA-Net consistently outperforms the baseline across both evaluation metrics.
        The axis definitions are the same as those in Fig.~\ref{fig:higgs_hhhr}.}
    \label{fig:higgs_hhbr}
\end{figure}

As $\HH \to 4b$ events are more likely to be detected at the LHC, we trained one SPA-Net using \HH resolved targets and another using both \HH boosted and resolved boosted targets.
We then evaluate their Higgs boson reconstruction purity and efficiency.
The corresponding Higgs boson reconstruction results are shown in Figs.~\ref{fig:higgs_hhr} and~\ref{fig:higgs_hhbr}, respectively.

Similar to the \HHH results, the \HH results demonstrate that SPA-Net achieves higher Higgs boson reconstruction purity and efficiency than the baseline, modified to target two reconstructed Higgs bosons with the first four jets ranked in Section~\ref{sec:baseline}, across all \pt bins.
Furthermore, the comparison between the resolved SPA-Net and the boosted+resolved SPA-Net shows that incorporating boosted Higgs boson targets can enhance SPA-Net's Higgs boson reconstruction purity and efficiency for $\pt > 400\GeV$.

\begin{figure}[htpb]
    \centering
    \includegraphics[width=0.45\textwidth]{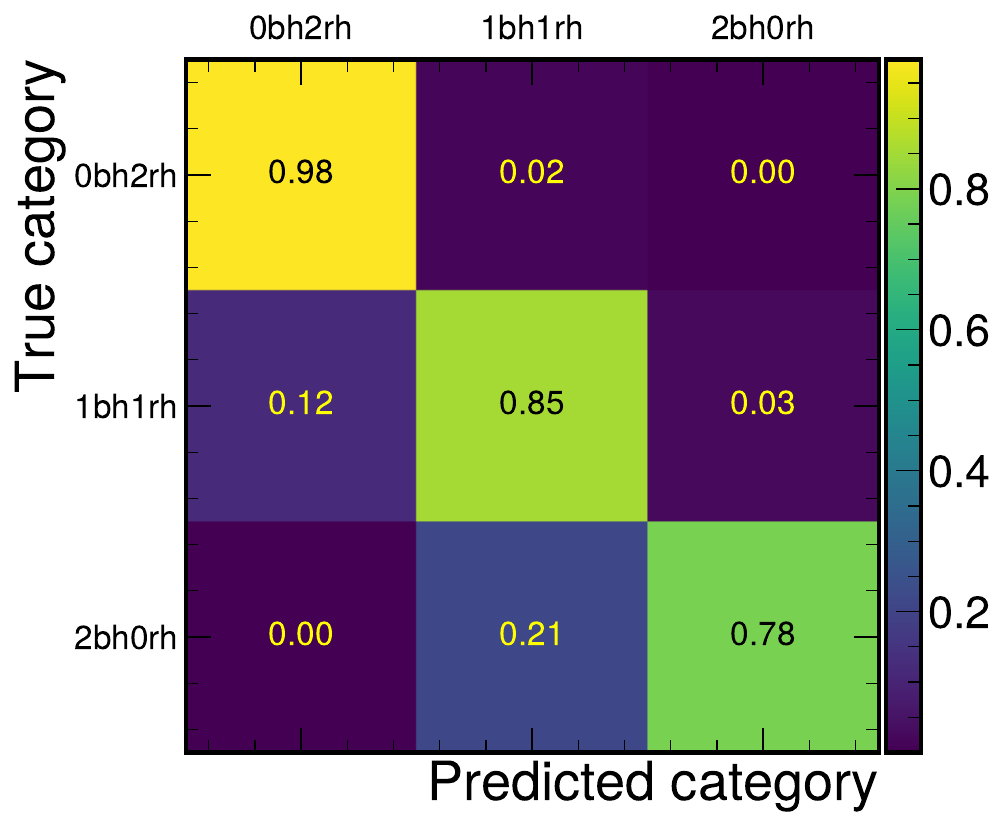}
    \caption{Confusion matrix for full events in the testing fraction of the \HH $m_\PH = 125\GeV$ dataset.
    Compared to Fig.~\ref{fig:hhh_confusion_matrix}, all true categories are maximized along the diagonal, indicating strong event topology classification performance.
    }
    \label{fig:hh_confusion_matrix}
\end{figure}

Similar to the confusion matrix in Fig.~\ref{fig:hhh_confusion_matrix}, Fig.~\ref{fig:hh_confusion_matrix} presents the confusion matrix for full events in the \HH dataset, illustrating that SPA-Net, combined with our postprocessing method, accurately classifies each target category based on the number of resolved and boosted Higgs boson candidates.


\section{Summary}
\label{sec:summary}

We generalized the symmetry-preserving attention network (SPA-Net) to reconstruct multiple Higgs bosons in \HHH and \HH events in both resolved and boosted topologies simultaneously, improving reconstruction performance in both.
Our \HHH boosted+resolved SPA-Net training demonstrated superior Higgs boson reconstruction purity and efficiency compared to a baseline approach combining a $\chi^2$-based methods and a boosted decision tree (BDT) large-radius jet tagger, as shown in Table~\ref{table:scalar_table}.
In particular, for \HHH (\HH) events considering both boosted and resolved topologies, our SPA-Net approach increases the \PH purity by 56\% (80\%) and the \PH efficiency by 38\% (37\%) compared to the baseline method.

\begin{table}[htpb]
    \caption{
        Summary of overall Higgs boson reconstruction purity and efficiency for the models discussed in Section~\ref{sec:results}.
        The row ``Reco. Target $\PH$" indicates the number of target Higgs bosons successfully recovered by each model.
        Most models show a decline in Reco. Target \PH when transitioning from the resolved dataset to the boosted+resolved dataset, except for the \HHH SPA-Net.
        This decline is attributed to the overlap removal described in Section~\ref{sec:postprocessing}, where more resolved jet assignments are removed than the number of additional boosted jet assignments included.
    }
    \resizebox{\linewidth}{!}{
        \begin{tabular}{l|cccccccc}
            \multirow{3}{*}{\textbf{Metrics}} & \multicolumn{4}{c|}{\HHH}    & \multicolumn{4}{c}{\HH}                                                                                                                                                                                                                 \\\cline{2-9}
                                              & \multicolumn{2}{c}{Resolved} & \multicolumn{2}{|c|}{Boosted+Resolved} & \multicolumn{2}{c}{Resolved}  & \multicolumn{2}{|c}{Boosted+Resolved}                                                                                                                          \\
                                              & Baseline                     & SPA-Net                                & \multicolumn{1}{|c}{Baseline} & SPA-Net                               & \multicolumn{1}{|c}{Baseline} & SPA-Net                              & \multicolumn{1}{|c}{Baseline} & SPA-Net         \\ \hline
            \PH Purity                        & 0.251                        & \multicolumn{1}{c|}{\textbf{0.444}}    & 0.261                         & \multicolumn{1}{c|}{\textbf{0.409}}   & 0.390                         & \multicolumn{1}{c|}{\textbf{0.702}}  & 0.402                         & \textbf{0.719}  \\
            \PH Efficiency                    & 0.394                        & \multicolumn{1}{c|}{\textbf{0.464}}    & 0.367                         & \multicolumn{1}{c|}{\textbf{0.507}}   & 0.578                         & \multicolumn{1}{c|}{\textbf{0.769}}  & 0.547                         & \textbf{0.752}  \\
            Reco. Target \PH                  & 32,223                       & \multicolumn{1}{c|}{\textbf{37,922}}   & 30,787                        & \multicolumn{1}{c|}{\textbf{42,486}}  & 47,039                        & \multicolumn{1}{c|}{\textbf{62,562}} & 45,481                        & \textbf{62,522} \\
            Target \PH                        & 81,810                       & \multicolumn{1}{c|}{81,810}            & 83,858                        & \multicolumn{1}{c|}{83,858}           & 81,330                        & \multicolumn{1}{c|}{81,330}          & 83,178                        & 83,178          \\
        \end{tabular}
    }
    \label{table:scalar_table}
\end{table}

To assess mass sculpting effects, we applied the boosted+resolved SPA-Net to a background multijet dataset and observed a less pronounced peak-like distribution compared to the baseline.
Furthermore, we extended the boosted+resolved SPA-Net to \HH events, where it similarly outperformed the $\chi^2$+BDT baseline.
Notably, in the high-\pt region, incorporating boosted targets in the training led to improved reconstruction purity and efficiency.

In addition, we proposed using predictions of SPA-Net to construct a likelihood for the number of reconstructed boosted and resolved Higgs bosons.
By further prioritizing boosted Higgs boson candidates and removing the overlapping resolved Higgs boson candidates, we can uniquely categorize events into boosted or resolved topologies.
We demonstrate that SPA-Net accurately categorizes events in most cases for both \HH and \HHH events.

Overall, our results show that the boosted+resolved SPA-Net algorithm provides significant improvements in Higgs boson reconstruction across both \HHH and \HH events.
Given the importance of maximizing the signal acceptance across different topologies to enhance the sensitivity to the small standard model cross sections, this method is a powerful tool for analyzing rare multi-Higgs boson final states at current and future colliders.

\acknowledgments

This work was supported by the Research Corporation for Science Advancement (RCSA) under grant \#CS-CSA-2023-109, Alfred P. Sloan Foundation under grant \#FG-2023-20452, U.S. Department of Energy (DOE), Office of Science, Office of High Energy Physics Early Career Research program under Award No. DE-SC0021187, and the U.S. National Science Foundation (NSF) Harnessing the Data Revolution (HDR) Institute for Accelerating AI Algorithms for Data Driven Discovery (A3D3) under Cooperative Agreement PHY-2117997.
This work was performed using the Pacific Research Platform Nautilus HyperCluster supported by NSF awards CNS-1730158, ACI-1540112, ACI-1541349, OAC-1826967, the University of California Office of the President, and the University of California San Diego's California Institute for Telecommunications and Information Technology/Qualcomm Institute.
Thanks to CENIC for the 100\,Gpbs networks.
Part of this research was conducted using computational resources and services at the Center for Computation and Visualization, Brown University.

\appendix

\section{Effect of \texorpdfstring{\PQb}{b}-Tagging Efficiency on Reconstruction Performance}
\label{sec:btagging}
\begin{figure}[htpb]
    \centering
    \includegraphics[width=0.5\textwidth]{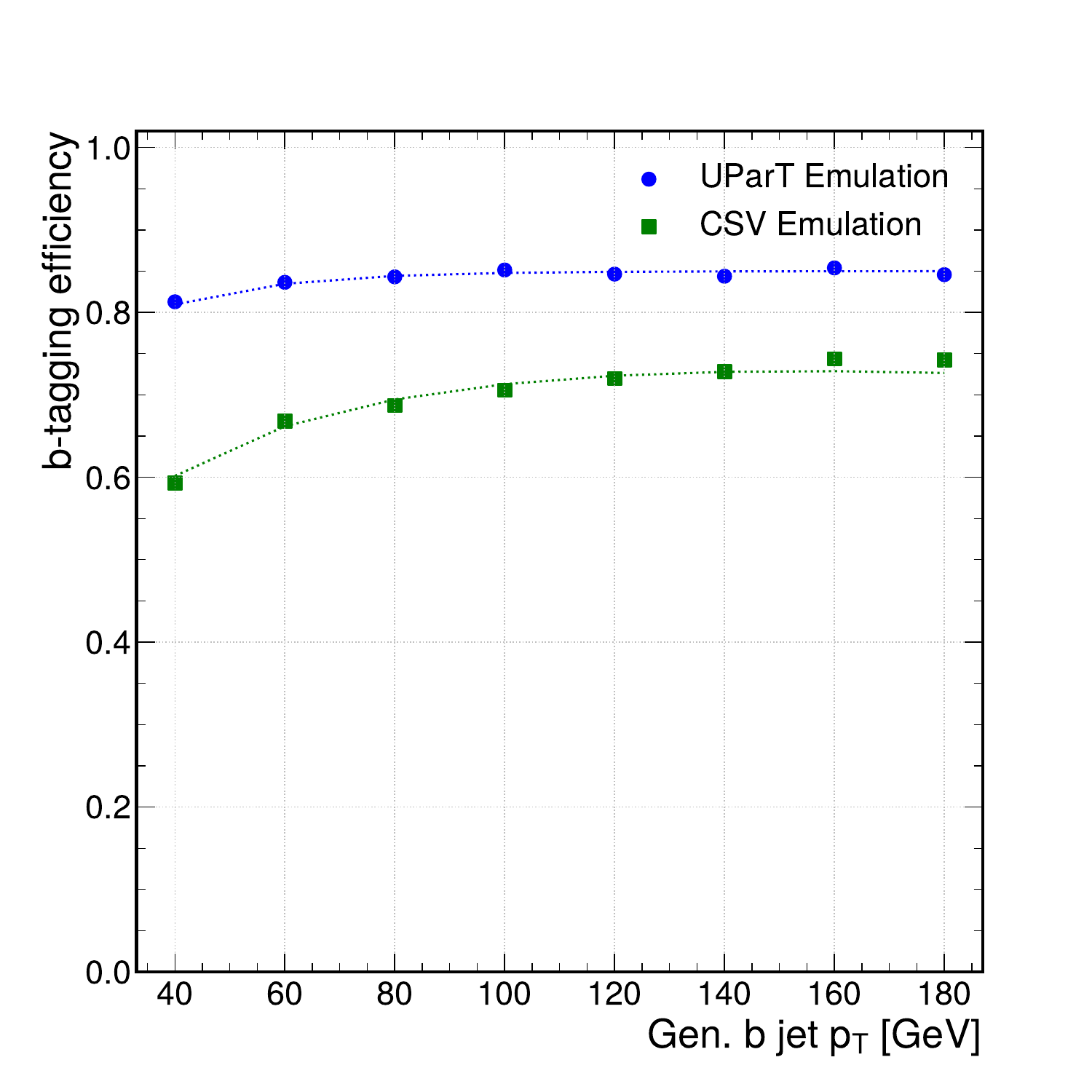}
    \caption{
    The parametrized \PQb-tagging efficiencies of the state-of-the-art Unified Particle Transformer (UParT) and the Combined Secondary Vertex (CSV) algorithms.
    Markers indicate efficiencies calculated directly from the dataset, while the dotted lines represent the parameterized equations.
    The \PQb-tagging efficiency improves from approximately 70\% to 85\%.
    }
    \label{fig:update_btagging}
\end{figure}

\begin{figure}[htpb]
    \centering
    \includegraphics[width=1\textwidth]{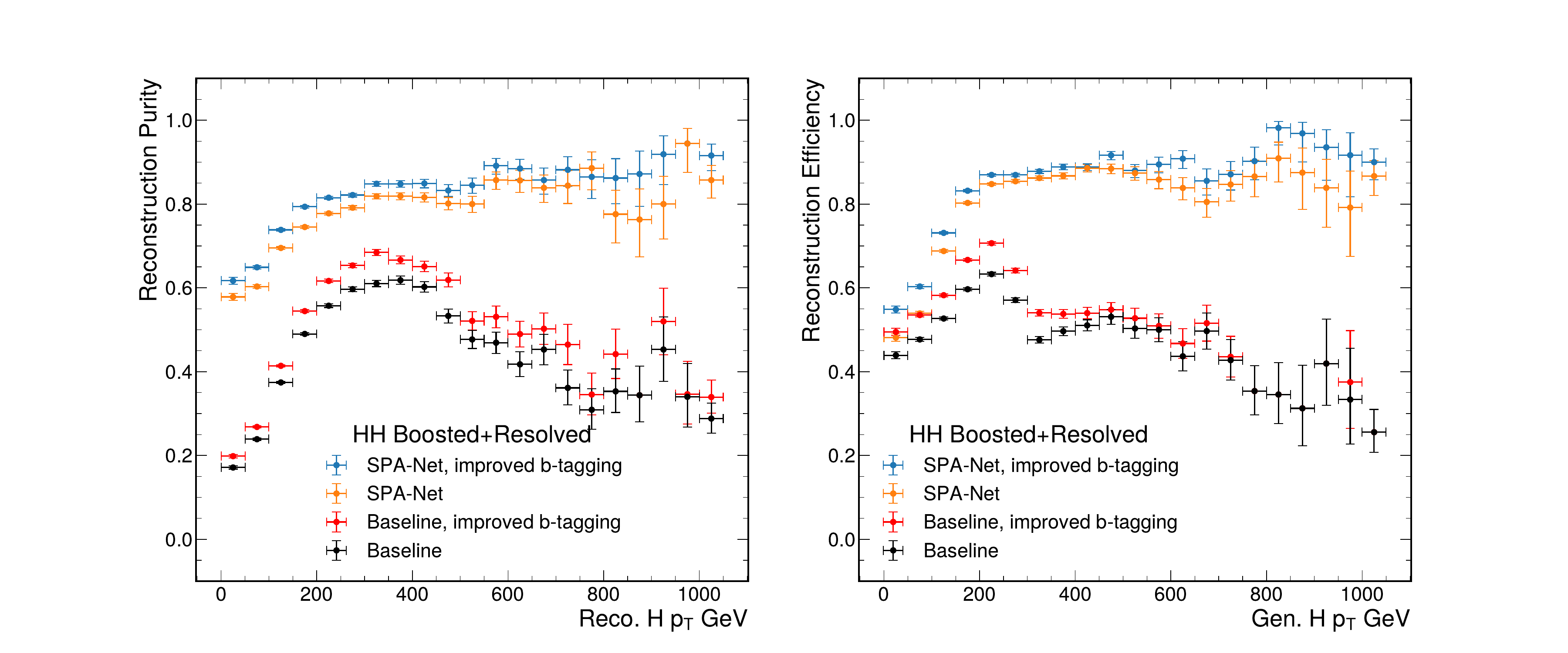}
    \caption{
     Reconstruction purity (left) and efficiency (right) for the baseline and SPA-Net, shown for both the conservative \PQb tagging (about 70\% efficiency) and updated \PQb tagging (about 85\% efficiency).
     Both methods benefit similarly from the improved \PQb tagging, indicating that the relative performance gain of SPA-Net remains stable across different \PQb-tagging configurations.
    }
    \label{fig:rerun_btagging}
\end{figure}

We further evaluated the impact of using an improved \PQb-tagging performance by emulating updated \PQb-tagging efficiencies based on recent experimental results~\cite{CMS-DP-2024-066}.
As shown in Fig.~\ref{fig:update_btagging}, the improved \PQb-tagging efficiency is about 85\% instead of 70\%, as used in previous sections.
This updated \PQb-tagging score was applied to both the $\chi^2$+BDT baseline method and as input to SPA-Net.
We observe that the reconstruction efficiency and candidate purity improved for both methods under the updated \PQb-tagging configuration, as shown in Fig.~\ref{fig:rerun_btagging}.
Notably, the relative improvements were comparable, indicating that the gains from enhanced \PQb tagging are similar across both approaches and do not qualitatively affect the conclusions drawn from our comparison.

\clearpage

\bibliographystyle{JHEP}
\bibliography{biblio}
\end{document}